\documentclass[sigconf,10pt]{acmart}
\usepackage{graphicx}
\usepackage[show]{chato-notes}
\usepackage{amsmath,amssymb}
\usepackage{epsfig}
\usepackage{algorithm}
\usepackage[noend]{algorithmic}

\usepackage{tikz}
\usetikzlibrary{shadows}
\usepackage{xcolor}
\usepackage[framemethod=TikZ]{mdframed}
\usepackage{tkz-berge,float}

\usepackage{url}
\usepackage{hyperref}
\hypersetup{
    bookmarks=true,         
    unicode=false,          
    pdftoolbar=true,        
    pdfmenubar=true,        
    pdffitwindow=false,     
    pdfstartview={FitH},    
    pdftitle={My title},    
    pdfauthor={Author},     
    pdfsubject={Subject},   
    pdfcreator={Creator},   
    pdfproducer={Producer}, 
    pdfnewwindow=true,      
    colorlinks=true,       
    linkcolor=black,          
    citecolor=black,        
    filecolor=black,      
    urlcolor=black           
}

\newcommand{\spara}[1]{\smallskip\noindent{\bf #1}}

\newtheorem{mydefinition}{Definition}
\newtheorem{myproperty}{Property}
\newtheorem{mytheorem}{Theorem}

\newtheorem{myexample}{Example}

\newtheorem{problem}{Problem}

\newtheorem{observation}{Observation}

\DeclareMathOperator*{\argmax}{arg\,max}
\newcommand{\bigO}{\mathcal{O}}

\newcommand{\NP}{$\mathbf{NP}$}
\newcommand{\NPhard}{$\mathbf{NP}$-hard}

\newcommand{\squishlist}{
 \begin{list}{$\bullet$}
  {  \setlength{\itemsep}{0pt}
     \setlength{\parsep}{3pt}
     \setlength{\topsep}{3pt}
     \setlength{\partopsep}{0pt}
     \setlength{\leftmargin}{2em}
     \setlength{\labelwidth}{1.5em}
     \setlength{\labelsep}{0.5em}
} }
\newcommand{\squishlisttight}{
 \begin{list}{$\bullet$}
  { \setlength{\itemsep}{0pt}
    \setlength{\parsep}{0pt}
    \setlength{\topsep}{0pt}
    \setlength{\partopsep}{0pt}
    \setlength{\leftmargin}{2em}
    \setlength{\labelwidth}{1.5em}
    \setlength{\labelsep}{0.5em}
} }

\newcommand{\squishdesc}{
 \begin{list}{}
  {  \setlength{\itemsep}{0pt}
     \setlength{\parsep}{3pt}
     \setlength{\topsep}{3pt}
     \setlength{\partopsep}{0pt}
     \setlength{\leftmargin}{1em}
     \setlength{\labelwidth}{1.5em}
     \setlength{\labelsep}{0.5em}
} }

\newcommand{\squishend}{
  \end{list}
}

\copyrightyear{2019}
\acmYear{2019}
\setcopyright{acmlicensed}
\acmConference[SIGMOD '19]{2019 International Conference on Management of Data}{June 30-July 5, 2019}{Amsterdam, Netherlands}
\acmBooktitle{2019 International Conference on Management of Data (SIGMOD '19), June 30-July 5, 2019, Amsterdam, Netherlands}
\acmPrice{15.00}
\acmDOI{10.1145/3299869.3324962}
\acmISBN{978-1-4503-5643-5/19/06}

\begin{document}

\title{Distance-generalized Core Decomposition}

\author{Francesco Bonchi}
\affiliation{%
	\institution{ISI Found. (Italy) \& Eurecat (Spain)}
}
\email{francesco.bonchi@isi.it}
\author{Arijit Khan}
\affiliation{%
	\institution{NTU, Singapore}
}
\email{arijit.khan@ntu.edu.sg}
\author{Lorenzo Severini}
\affiliation{%
	\institution{ISI Foundation, Turin, Italy}
}
\email{lorenzo.severini@isi.it}

\renewcommand{\shortauthors}{F. Bonchi et al.}

\begin{abstract}
The $k$-core of a graph is defined as the maximal subgraph in which
every vertex is connected to at least $k$ other vertices within that
subgraph. In this work we introduce a distance-based generalization of the
notion of $k$-core, which we refer to as the $(k,h)$-core, i.e.,
the maximal subgraph in which every vertex has at least $k$ other vertices
at distance $\leq h$ within that subgraph.
We study the properties of the $(k,h)$-core showing that it preserves many of the nice features of the classic core decomposition (e.g., its connection with the notion of \emph{distance-generalized chromatic number}) and it preserves its usefulness to speed-up or approximate distance-generalized notions of dense structures, such as \emph{$h$-club}.

Computing the distance-generalized core decomposition over large networks is intrinsically complex. However, by exploiting clever upper and lower bounds we can partition the computation in a set of totally independent subcomputations, opening the door to top-down exploration and to multithreading, and thus  achieving an efficient algorithm.

\end{abstract}
\maketitle \sloppy

\section{Introduction}
\label{sec:intro}
Extracting dense structures from large graphs has emerged as a key graph-mining primitive in a variety of application scenarios~\cite{aggarwal}, ranging from web mining~\cite{gibson}, to biology~\cite{fratkin}, and finance~\cite{Du}. Many different definitions of dense subgraphs have been proposed (e.g., \emph{cliques, n-cliques, n-clans, k-plexes, f-groups, n-clubs, lambda sets}), but most of them are
\NPhard\ or at least quadratic. In this respect, the concept of \emph{core decomposition} is particularly
appealing because ($i$) it can be computed in linear time~\cite{MatulaB83,BatageljFastCores11}, and ($ii$) it is related to many of the various definitions of a dense subgraph and it can be used to speed-up or approximate their computation.

The $k$-\emph{core} of a  graph is defined as a maximal subgraph in which every vertex is connected to at least $k$ other vertices within that subgraph.
The set of all $k$-cores of a graph, for each $k$, forms its \emph{core decomposition}~\cite{Seidman1983k-cores}.
The \emph{core index} of a vertex $v$ is the maximal $k$ for which $v$ belongs to the $k$-core.

While core decomposition is based on the number of immediate connections that a vertex has within a subgraph (its degree), the importance of studying network structures beyond the horizon of the distance-1 neighborhood of a vertex is well established since several decades, especially in social sciences~\cite{L50}.
Following this observation, in this paper we introduce an important generalization of the notion of core decomposition. Looking through the lens of \emph{shortest-path distance}, one can see the degree of a vertex $v$ as the number of vertices in the graph which have distance $\leq 1$ from $v$, or equivalently, the size of the 1-neighborhood of $v$. From this perspective, a natural generalization is to consider a distance threshold $h > 1$. This leads smoothly to the notions of \emph{$h$-neighborhood}, \emph{$h$-degree}, and in turn,  to the distance-generalized notion of $(k,h)$-core, i.e., the maximal subgraph in which every vertex has at least $k$ other vertices at distance $\leq h$ within that subgraph.
As we formally prove later, the $(k,h)$-core is unique and it is contained in the $(k-1,h)$-core: these facts allow us to define the notion of \emph{distance-generalized core decomposition}.

\begin{figure}[t!]
\vspace{2mm}
  \centering
  \includegraphics[width=\columnwidth]{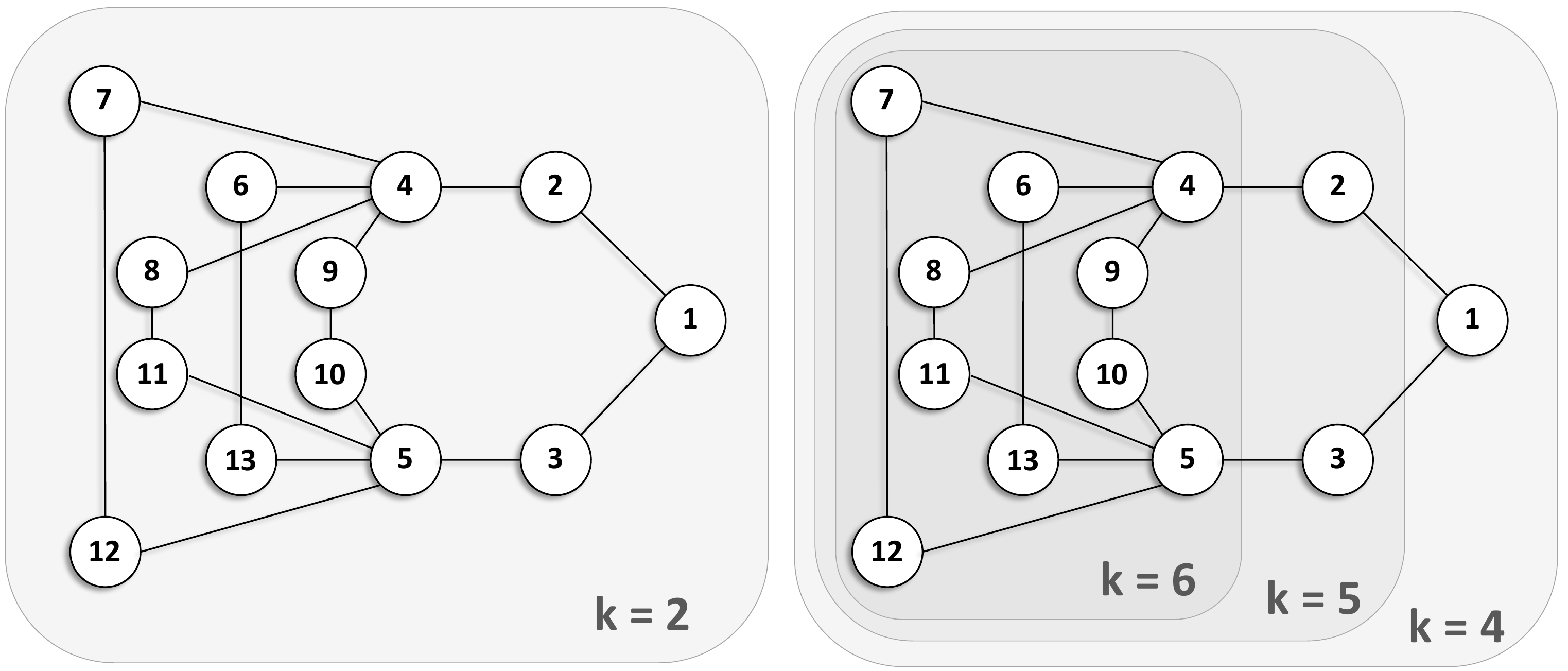}

 \vspace{-2mm}

  \caption{On the left-hand side, the $(k,1)$-core decomposition of an example graph (i.e., the classic core decomposition). On the right-hand side, the $(k,2)$-core decomposition of the same graph.}\label{fig:firstexemple}
 \end{figure}
\begin{myexample}
Figure \ref{fig:firstexemple} shows the differences between the classic core decomposition (on the left side) and the $(k,2)$-core decomposition (on the right side). For this example graph, the classic core decomposition (i.e., the  $(k,1)$-core decomposition in our setting) puts all the vertices in core $k=2$. On the contrary, by considering distance-2 neighborhood, the $(k,2)$-core decomposition is able to capture structural differences among the vertices, thus providing a finer-grained analysis. In particular,  it allows detecting the fact that the vertices from 4 to 13 form a denser and more structured region (the  $(6,2)$-core), while vertices 2 and 3 are in the  $(5,2)$-core, and vertex 1 only belongs to the $(4,2)$-core.
\end{myexample}

\spara{Challenges and contributions.}
 In this paper we show that: $(1)$ the introduced notion of distance-generalized core decomposition is useful in practice, $(2)$ its computation is much harder than the classic core decomposition, and $(3)$ nevertheless, efficiency and scalability can be achieved.

For what concerns $(1)$, we show  that distance-generalized core decomposition generalizes many of the nice properties of the classic core decomposition, e.g., its connection with the notion of \emph{distance-generalized chromatic number}, or  its usefulness in speeding-up or approximating distance-generalized notions of dense structures, such as \emph{$h$-club} and \emph{(distance-generalized) densest subgraph}. We also show that it is very informative and performs well as a heuristic for \emph{selecting landmarks} to build shortest-path-distance oracles.

As it happens for the distance generalization of other notions (an example is the \emph{maximum $h$-club problem} that we discuss in \S\ref{subsec:club} and \S\ref{subsec:exp3}), the computation of
the distance-generalized core decomposition is a daunting task.

\begin{figure}[t!]
\vspace{-2mm}
  \centering
 \includegraphics[width=.7\columnwidth]{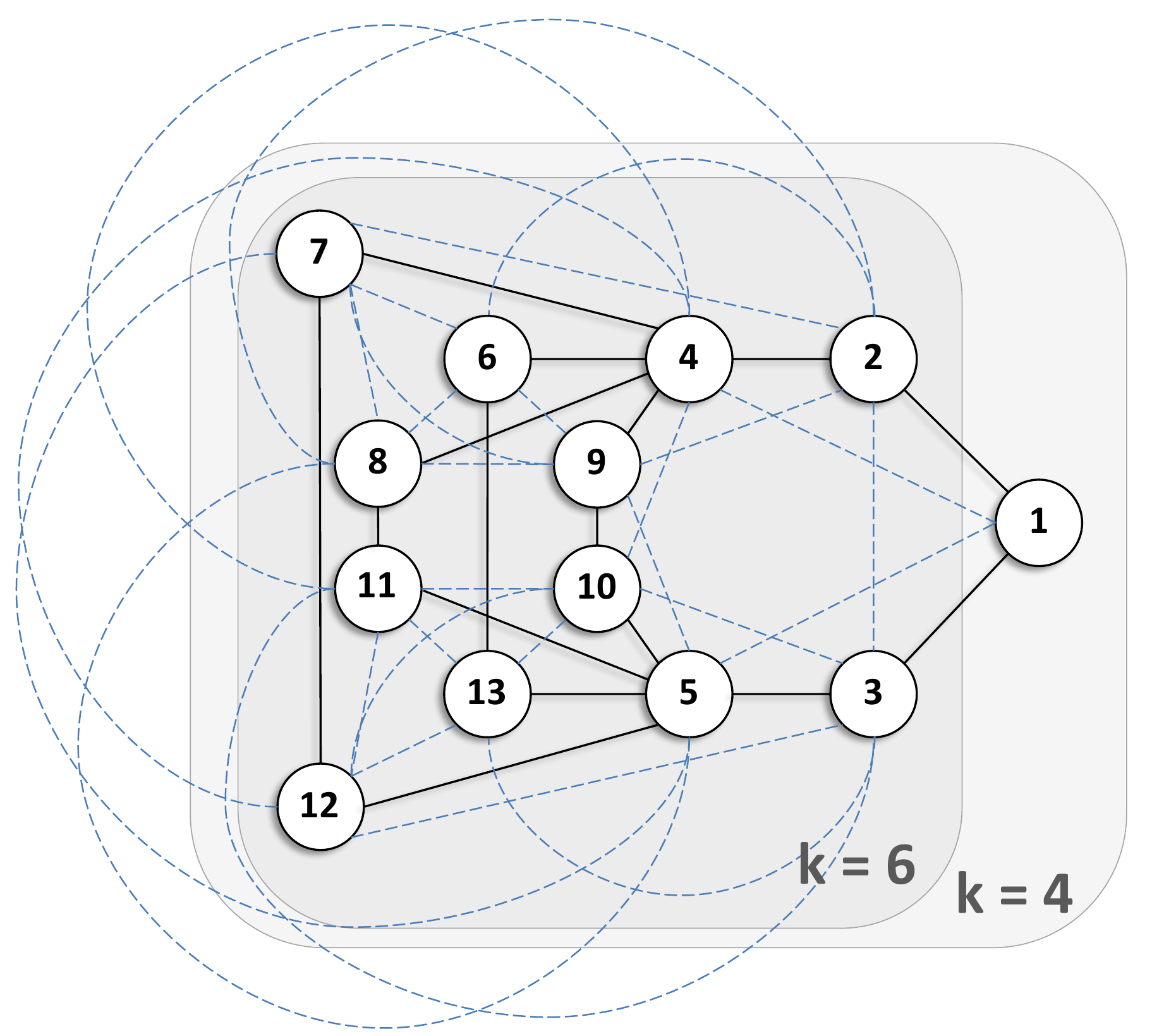}
 \vspace{-4mm}

  \caption{The power-graph $G^h$ of the graph in Figure~\ref{fig:firstexemple} for $h= 2$. The dotted edges are added between pairs of vertices at distance = 2. It can be observed that the core decomposition of $G^2$ does not correspond to the $(k,2)$-core decomposition of $G$ (rightside of  Figure \ref{fig:firstexemple}). }\label{fig:secondexemple}
\vspace{-4mm}
 \end{figure}

One could think to obtain the distance-generalized  $(k,h)$-core decomposition,
by first computing the $h$-power\footnote{\scriptsize The $h$-power $G^h$ of an undirected graph $G$ is another graph that has the same set of vertices, but in which two vertices are adjacent when their distance in $G$ is at most $h$. See \url{https://en.wikipedia.org/wiki/Graph_power}}  of the input graph (as in Figure \ref{fig:secondexemple}), and then applying the state-of-the-art algorithms for core decomposition. However, this does not provide the correct decomposition, as shown next.

\begin{myexample} Figure \ref{fig:secondexemple} shows the power-graph $G^2$ of the graph in Figure \ref{fig:firstexemple}. We can observe that according to the classic core decomposition of $G^2$, the vertices 2 and 3 have core-index 6, while in the $(k,2)$-core decomposition of $G$ (rightside of  Figure \ref{fig:firstexemple}) they have core-index 5. This is due to the fact that in $G^2$, vertices 2 and 3 becomes adjacent due to vertex 1, but this vertex does not belong to the $(5,2)$-core.
\end{myexample}
In the classic core decomposition, when a vertex is removed, the degree of its neighbors is decreased by 1: this observation allows several optimizations which makes the classic case easy to compute and to scale \cite{DistributedCores1,DistributedCores2,StreamingCores,cheng2011efficient,Khaouid2015,BatageljFastCores11}. Instead, in the generalized $(k,h)$-core decomposition the removal of a vertex can decrease the $h$-degree of some of its $h$-neighbors by more than one. This is the reason why the idea of decomposing the $h$-power graph does not work, and\emph{ it is also the main reason why distance-generalized core decomposition ($h > 1$) is much harder to compute than standard core decomposition ($h = 1$)}.  In fact, when a vertex is removed, we need to compute again the $h$-degree of a large number of vertices, i.e., we need a
large number of $h$-bounded BFS traversals. In spite of such challenges, we devise efficient algorithms.
As a baseline, we first extend the state-of-the-art Batagelj and Zaver{\v{s}}nik's algorithm \cite{BatageljFastCores11} to deal with $(k,h)$-core decomposition (we dub the algorithm $h$\textsf{-BZ}). Next, we exploit a \emph{lower bound} on the core-index of a vertex to avoid a large number of useless $h$-degree re-computations. We call this algorithm $h$\textsf{-LB}. Finally, we propose an algorithm that further improves efficiency by computing an \emph{upper bound} and processing the vertices with larger $h$-degrees as early as possible
(dubbed $h$\textsf{-LB+UB}). In order to scale to larger graphs we also exploit multi-threading provided by modern architectures to parallelize the computations of $h$-degrees.

\smallskip
The main contributions of this paper are as follows:
\squishlist
	\item We introduce the problem of distance-generalized core decomposition
	and characterize the properties of the $(k,h)$-core of a given network (\S\ref{sec:problem}).

	\item We generalize a state-of-the-art algorithm for core decomposition, to deal with our problem. Then we propose two exact and efficient algorithms equipped with lower and upper-bounding techniques (\S\ref{sec:algorithms}).
	
	\item We prove a connection between distance-generalized core decomposition and \emph{distance-generalized chromatic number} (\S\ref{subsec:chromatic}).
We prove that every $h$-club of size $k+1$ must be included in the $(k,h)$-core, and exploit this property to develop an efficient algorithm for the hard problem of \emph{maximum $h$-club} (\S\ref{subsec:club}). We introduce the novel problem of \emph{distance-generalized densest subgraph}, and prove that by using our $(k,h)$-core decomposition, we can obtain an efficient algorithm with approximation guarantees  (\S\ref{subsec:densest}).


\item Our thorough experimentation (\S\ref{sec:experiments}) confirms the effectiveness of the proposed bounds in enhancing the efficiency of our algorithms.
 Experiments on the \emph{maximum $h$-club} problem demonstrate that our proposal of using  the distance-generalized core decomposition as a pre-processing achieves a \emph{consistent speed-up over the state-of-the-art methods for this hard problem} (\S\ref{subsec:exp3}).

%

%
\squishend

For space economy, all proofs are given in the Appendix, which also contains an additional application and additional experiments.  

\section{Background and related work}
\label{sec:related}

\spara{Core decomposition.} Consider an undirected, unweighted  graph $G=(V,E)$, where $V$ is the set of vertices and $E \subseteq V \times V$
is the set of edges. Given a subset of vertices $S \subset V$, we denote $G[S] = (S,E[S])$ the subgraph of $G$ induced by $S$, i.e., $E[S] =  \{(u,v) \in E | u, v \in S\}$. For each vertex $v \in V$, let $deg_G(v)$ denote the degree of $u$ in $G$.
\begin{mydefinition}[core decomposition]\label{def:kcores}
  The $k$\emph{-core} of a graph $G=(V,E)$ is a
  \emph{maximal} subgraph $G[C_k] = (C_k,E[C_k])$ such that $\forall v \in C_k: deg_{G[C_k]}(v) \geq k$.
  The set of all $k$-cores $V = C_0 \supseteq C_1 \supseteq \cdots \supseteq C_{k^*}$ ($k^* = \arg\max_{k} C_k \neq \emptyset$) is the
  \emph{core decomposition} of $G$. We identify a core either with the set $C_k$ of vertices or with its induced subgraph $G[C_k]$, indifferently.
\end{mydefinition}

Core decomposition can be computed
by iteratively removing the smallest-degree vertex and setting its core number as its degree at the time of removal.
It can be used to speed-up the computation of \emph{maximal cliques}~\cite{EppsteinLS10}, as a
clique of size $k$ is guaranteed to be in a $(k\!-\!1)$-core, which can be
significantly smaller than the original graph.
 Moreover, core decomposition is at the basis of linear-time approximation algorithms for the \emph{densest-subgraph problem} \cite{KortsarzP94} and the densest at-least-$k$-subgraph problem \cite{AndersenC09}.
It is also used to approximate betweenness centrality~\cite{HealyJMA06}. Furthermore, the notion of core index is related to \emph{graph coloring} and to the notion of \emph{chromatic number} \cite{SZEKERES19681,MatulaB83}. It has been employed for analyzing/visualizing complex networks \cite{DBLP:conf/gd/BatageljMZ99,Alvarez-HamelinDBV05}
in several domains, e.g., bioinformatics \cite{DBLP:journals/bmcbi/BaderH03,citeulike:298147}, software engineering~\cite{DBLP:journals/tjs/ZhangZCLZ10},
and social networks~\cite{Kitsak2010,GArcia2013}.
It has been studied under various settings, such as distributed \cite{DistributedCores1,DistributedCores2}, parallel~\cite{Dhulipala2017}, streaming \cite{StreamingCores}, and disk-based \cite{cheng2011efficient}, and for various types of graphs, such as uncertain \cite{bonchi14cores}, directed \cite{DirectedCores}, multilayer \cite{galimberti17}, temporal \cite{galimberti18}, and weighted \cite{WeigthedCores} graphs.

\spara{Generalizations and variants.} Several generalizations and variants of the concept of core decomposition have been proposed recently. A popular one is \emph{$k$-truss} \cite{wang2012truss,ZhangP12,ZhaoT12,KtrussSIGMOD14} defined as a subgraph where any edge is involved in at least $k$ triangles.
Sariyuce  et  al.  \cite{SariyuceSPC15} introduce the notion of \emph{nucleus decompositions}
which generalizes $k$-core by defining it
on a subgraph. Sariyuce and Pinar~\cite{SariyuceP16} develop generic algorithms to construct the hierarchy of dense subgraphs (for
$k$-core,
$k$-truss, or any nucleus decomposition) in such a way to keep track of the different
connected  components.

Tatti and Gionis \cite{TattiG15} define \emph{density-friendly graph decomposition}, where  the  density  of  the  inner  subgraphs  is  higher than the density of the outer subgraphs.
Govindan et al. \cite{GovindanWXDS17} propose \emph{$k$-peak decomposition} which allows finding local dense regions in contrast with classic core decomposition. 
Zhang et al.~\cite{Zhang2017} propose the $(k,r)$-core decomposition considering both user engagement and similarity.

\spara{Distance-generalization of cliques.} Several distance-based generalizations of the notion of clique
have been proposed \cite{alba1973,Balasundaram2005,L50,M79}. A subset of vertices $S \subseteq V$ is an {\em $h$-clique} if $d(u,v)\le h$ for all $u,v \in S$ (where $d(u,v)$ denotes the shortest-path distance between $u$ and $v$).  Note that an $h$-clique $S$ may be a
disconnected set of vertices, as vertices outside of $S$ might be used to define the shortest-path distance between two vertices in $S$.
To avoid such problem, the concept of  $h$-club was defined as a subset of vertices $S \subseteq V$ whose induced subgraph $G[S]$ has diameter at most $h$.

An $h$-clique (or $h$-club) is said to be \emph{maximal} when it is maximal by set inclusion, i.e., it is not a proper subset of a larger $h$-clique (or $h$-club respectively).
An $h$-clique (or $h$-club) is said to be \emph{maximum} it there is no larger $h$-clique (or $h$-club, respectively) in $G$.
The problems of finding an $h$-clique or an $h$-club of maximum cardinality are both  \NP-hard, for any fixed positive integer $h$; and they
remain hard even in graphs of fixed diameter \cite{Balasundaram2005}. In contrast to
maximum $h$-clique, the \emph{maximum $h$-club problem} is even more complex due to the fact that $h$-clubs are not closed under set inclusion: i.e., a subset of a $h$-club may not be an $h$-club. Indeed, even only testing inclusion-wise maximality of $h$-clubs is \NP-hard~\cite{MAHDAVIPAJOUH12}.
Several exact and approximated methods have been developed for the maximum $h$-club problem \cite{AC12,Balasundaram2005,CHLS13,BLP00,SKMN12,SB13,VPP15,VEREMYEV2012316,MB15}.
In this paper (\S\ref{subsec:club}), we show how to exploit distance-generalized core decomposition to speed-up these existing methods.

\spara{Densest subgraph.}
Among the various notions of a dense subgraph, the problem of extracting the subgraph maximizing
the \emph{average-degree density} (a.k.a. the \emph{densest subgraph}) has attracted most of the research in the field, because it is solvable in polynomial time \cite{Goldberg84}, and it has
a fast $\frac{1}{2}$-approximation algorithm~\cite{AITT00,Char00}, which resembles the algorithm for core decomposition: greedily remove the smallest-degree vertex until the graph is emptied, then return the densest among all subgraphs produced during this vertex-removal process.


In this paper (\S\ref{subsec:densest}) we introduce the notion of \emph{distance-generalized densest subgraph} as the subgraph that maximizes the average $h$-degree (for a given $h \geq 1$) of the vertices in the subgraph. We show that, similar to the classic densest subgraph, by means of the distance-generalized core decomposition we can achieve a fast algorithm with approximation guarantees for this novel problem.

\section{Problem statement}
\label{sec:problem}
Given an undirected, unweighed graph $G=(V,E)$ and a set of vertices $S \subseteq V$, let $G[S] = (S,E[S])$ denote the subgraph induced by $S$.
Given a positive integer $h \in \mathbb{N}^+$, the $h$-neighborhood in $G[S]$ of a vertex $v \in S$ is defined as $N_{G[S]}(v,h) = \{u \in S | u \neq v, d_{G[S]}(u,v) \leq h \}$ where $d_{G[S]}(u,v)$ is the shortest-path distance between $u$ and $v$ computed over $G[S]$, i.e., using only edges in $E[S]$. We also define the $h$-degree of a vertex as the size of its $h$-neighborhood: i.e., $deg^h_{G[S]}(v) = |N_{G[S]}(v,h)|$.
\begin{mydefinition}[$(k,h)$-core] Given a distance threshold $h \in \mathbb{N}^+$ and an integer $k \ge 0$, the $h$-neighborhood $k$-core ($(k,h)$-core for short) of a graph $G$ is a \emph{maximal} subgraph $G[C_k] = (C_k,E[C_k])$ such that $\forall v \in C_k: deg^h_{G[C_k]}(v) \geq k$.
\end{mydefinition}
The $(k,h)$-core has the following two properties.
\begin{myproperty}[Uniqueness]
\label{prop:unique}
Given a threshold distance  $h \in \mathbb{N}^+$ and an integer $k \ge 0$,
the $(k,h)$-core of a graph $G$ is unique.
\end{myproperty}
%
%
\begin{myproperty}[Containment]
\label{prop:subgraph}
Given a distance  threshold $h \in \mathbb{N}^+$ and an integer $k \ge 0$,
the $(k+1,h)$-core  of a graph $G$ is a subgraph of its $(k,h)$-core.
\end{myproperty}
%
%
Given that the $(k,h)$-core is unique and it is a supergraph of the $(k+1,h)$-core, as in the classic core decomposition it holds that $V = C_0 \supseteq C_1 \supseteq \cdots \supseteq C_{k^*}$ ($k^* = \arg\max_{k} C_k \neq \emptyset)$.  The fact that cores are all nested one into another as in the classic core decomposition also allows associating  to each vertex $v \in V$ a unique \emph{core index} (denoted $core(v)$), i.e., the maximum $k$ for which $v$ belongs to the $(k,h)$-core.  Not only does the situation resemble that of the classic core decomposition, it is straightforward to observe that our definition of $(k,h)$-core perfectly generalizes the classic notion of $k$-core: in fact for $h = 1$, the generalized notions of $h$-neighborhood, $h$-degree, and thus $(k,h)$-core correspond to the classic notions of neighborhood, degree, and $k$-core, respectively. While the problem for $h = 1$ has been widely studied (as discussed in \S\ref{sec:related}), the problem studied in this paper is how to compute efficiently, for a given distance threshold $h > 1$, the \emph{distance-generalized core decomposition}, i.e., the set of all the non-empty $(k,h)$-cores.


\section{Algorithms}
\label{sec:algorithms}
In this section we design three {\em exact} algorithms for computing the $(k,h)$-core decomposition, for a given distance threshold $h >1$.
First, we introduce $h$\textsf{-BZ}, the distance-generalized version
of the classic Batagelj and Zaver{\v{s}}nik's
method~\cite{BatageljFastCores11} (\S\ref{sec:peeling_algo}).
We then develop two more efficient algorithms: the first one ($h$\textsf{-LB}) exploiting a {\em lower bound} on the core index of a vertex (\S\ref{sec:lb:algo}), the second one ($h$\textsf{-LB+UB})
employing an {\em upper bound} on the core index, thereby computing the core index of high $h$-degree vertices as early as possible in a \emph{top-down} fashion
(\S\ref{sec:topdown_algo}).
\subsection{Baseline: Distance-generalized Batagelj and Zaver{\v{s}}nik's method}
\label{sec:peeling_algo}
Based on Property~\ref{prop:subgraph}, $(k+1,h)$-core can be obtained
 by ``peeling" the $(k,h)$-core. This means to recursively delete, from the $(k,h)$-core of $G$,  all the vertices with $h$-degree less than $k+1$: what remains is the $(k+1,h)$-core
of $G$.

Algorithm~\ref{alg:peeling_algo} processes the vertices
in increasing order of their $h$-degrees, using a vector $B$ of lists, where each cell $i$ in $B[i]$ is a list containing all vertices with $h$-degree equal to $i$. This technique of maintaining the vertices sorted during the computation is called \emph{bucketing}, and it allows updating each cell (or bucket) in $\bigO(1)$ time.\footnote{\scriptsize Recently, Khaouid et al.~\cite{Khaouid2015} propose an efficient implementation of the  {\sf B-Z} algorithm for classic core decomposition maintaining the vertices in a flat array of size  $|V|$. Adopting the same technique in our context would be inefficient as
the deletion of a single vertex can decrease the $h$-degree of its $h$-neighbors by more than 1. In a flat array, moving a vertex to a cell is linear to the distance between the old and the new cells, since it requires a linear number of swaps between intermediate values. Instead, we model $B$ as a vector of lists.} After initializing $B$ (Lines 1--3), Algorithm~\ref{alg:peeling_algo}  processes the cells of $B$ (and the vertices therein) in increasing order.
When a vertex $v$ is processed at iteration $k$, it is removed from the set of alive vertices $V$, its core index is set to $k$.
When we delete a vertex, the $h$-degree of the vertices in its $h$-neighborhood decreases, and these vertices are moved in the appropriate cell of $B$
(Lines 8--10).
The algorithm terminates when all vertices in $V$ are processed and their core indexes are computed.
\begin{algorithm}[tb!]
\caption{$h$\textsf{-BZ}}
\label{alg:peeling_algo}
\begin{algorithmic}[1]
		\REQUIRE graph $G=(V,E)$, distance threshold $h >1$
		\ENSURE core index $core(v)$ for each vertex $v \in V$
		\FORALL{$v \in V$}
		\STATE compute $deg^h_{G[V]}(v)$
		\STATE $B[deg^h_{G[V]}(v)]\leftarrow B[deg^h_{G[V]}(v)] \cup \{v\}$
		\ENDFOR
		\FORALL{$k=1,2,\ldots,|V|$}
		\WHILE{$B[k] \neq \emptyset$}
		\STATE pick and remove a vertex $v$ from $B[k]$
		\STATE $core(v) \leftarrow k$
		\FORALL{$u \in N_{G[V]}(v,h)$}
		\STATE compute $deg^h_{G[V\setminus\{v\}]}(u)$
		\STATE move u to $B[\max \left(deg^h_{G[V\setminus\{v\}]}(u), k\right)]$
		\ENDFOR
		\STATE $V \leftarrow V\setminus\{v\}$
		\ENDWHILE
		\ENDFOR
	\end{algorithmic}
\end{algorithm}

\smallskip

\noindent  \textbf{Correctness.} Let $\tilde{V}$ and $\tilde{k}$ denote the current status of $V$ and $k$, respectively. At the beginning of every iteration of the outer for-loop (lines 4--11), it holds that $u \in B[i] \implies  deg^h_{G[\tilde{V}]}(u) = i$. This is true at the initialization of $B$ (line 3). The $h$-degree of $u$ w.r.t. the current $\tilde{V}$ can only shrink when one of its $h$-neighbors is removed. When this happens (line 6), the $h$-degree of $u$ is recomputed (line 9) and $u$ is reassigned to a new bucket corresponding to $deg^h_{G[\tilde{V}]}(u)$ (line 10), until we find that the removal of an $h$-neighbor $v$ of $u$ shrinks the $h$-degree of $u$ below the current $\tilde{k}$. When this happens, it means that we have found the core index $\tilde{k}$ of $u$. In fact $deg^h_{G[\tilde{V}\cup \{v\}]}(u) \geq \tilde{k}$ (as $u$ is still alive)
and all its $h$-neighbors in $\tilde{V}\cup \{v\}$ have $h$-degree $\geq \tilde{k}$ w.r.t. $\tilde{V}\cup \{v\}$ (as they are still alive). Therefore $u \in (\tilde{k},h)$-core. However, the removal of $v$ decreases the $h$-degree of $u$ under $\tilde{k}$, so that when the value of $k$ will increase to $\tilde{k}+1$, $u$ will not have enough $h$-neighbors still alive, thus it cannot belong to the $(\tilde{k} + 1,h)$-core.

Finally, after removing $v$ and detecting that $deg^h_{G[\tilde{V}]}(u) < \tilde{k}$, the algorithm introduces $u$ in $B[\tilde{k}]$ (line 10). Future removals of $h$-neighbors of $u$ will maintain $u$ in $B[\tilde{k}]$ (line 10), until it comes the turn of $u$ to be picked from $B[\tilde{k}]$ (line 6), and its core index to be correctly assigned as $\tilde{k}$ (line 7).

\spara{Computational complexity.}
The time complexity of Algorithm~\ref{alg:peeling_algo} is
$\bigO(|V| D (D + \tilde{E}))$.
Here, $D$ and $\tilde{E}$ are the maximum size of the subgraph induced by an $h$-neighborhood of a vertex, in terms of the number of vertices (i.e., the $h$-degree) and edges, respectively.
This is because Algorithm~\ref{alg:peeling_algo} iterates over all vertices. While processing a vertex, we re-compute the $h$-degree of all vertices within its $h$-neighborhood, which requires $(D + \tilde{E})$ time for each vertex.

\subsection{Lower bound algorithm}
\label{sec:lb:algo}
The baseline $h$\textsf{-BZ} algorithm described above is inefficient over large and dense networks,
since, for every vertex deleted, it re-computes the $h$-degrees of {\em all} vertices
within its $h$-neighborhood. In this regard, we ask the following critical question:
is it necessary to re-compute
the $h$-degrees of {\em all} vertices in the $h$-neighborhood of a deleted vertex? In fact, suppose that we can know in advance a \emph{lower bound} on the value of the core index of a certain vertex. Then, we have the guarantee that such vertex will not be removed for values of $k$ smaller than its lower-bound,
so we can avoid to update its $h$-degree until the value of $k$ has reached its lower bound.

In the following we prove a natural lower bound $LB_1()$ on the core index of a vertex.
\begin{observation}
	\label{th:lb1}
$core(v)\geq deg^{{\lfloor\frac{h}{2}}\rfloor}_{G[V]}(v)=LB_1(v)$
\end{observation}
	The main idea of $LB_1$ is that every vertex in the $\lfloor\frac{h}{2}\rfloor$-neighborhood for $v$ has at least ${\lceil\frac{h}{2}\rceil}$ neighbors at distance $h$ (i.e., the vertex $v$ plus the other  $({\lfloor\frac{h}{2}\rfloor}-1)$-neighbors of $v$).
%
%
Next, we further refine the lower bound of $core(v)$ by considering the value of $LB_1$ of the ${\lceil\frac{h}{2}\rceil}$-neighbors of $v$.
\begin{observation}
\label{th:lb2}
\begin{align}
core(v)&\geq \max\left\lbrace \{LB_1(u) : d_{G[V]}(u,v)\leq {\lceil\frac{h}{2}\rceil}\},  LB_1(v) \right\rbrace & \nonumber \\
& = LB_2(v)& \nonumber
\end{align}
\end{observation}
%

Notice that $LB_2(v)$ improves $LB_1(v)$ by taking the largest $LB_1$ among the vertices in $({\lceil\frac{h}{2}\rceil})$-neighborhood (since every vertex at distance $\lfloor\frac{h}{2}\rfloor$ from ${\lceil\frac{h}{2}\rceil}$-neighborhood is at most at distance $h$ from $v$). The above observation does not hold for values greater than $\frac{h}{2}$ because in that case some vertex might be at a distance greater than $h$ from each other.

\begin{myexample}
	Consider the graph in Figure~\ref{fig:firstexemple}, and assume $h = 2$. We have $LB_1(v_1)= LB_1(v_2)=2$, and $LB_1(v_4)=5$ . Since $v_4$ is in the 1-neighborhood of $v_2$, we have that $LB_2(v_2)=\max \left( LB_1(v_2),LB_1(v_4)\right) $  $= 5 \leq core(v_2) = 5$.
\end{myexample}

Based on this lower bound, we devise the $h$\textsf{-LB}, whose pseudocode
is presented in Algorithms~\ref{alg:lb} and \ref{alg:partition}.
While the overall flow follows that of the baseline $h$\textsf{-BZ} algorithm, the use of the lower bound reduces
the number of $h$-degree re-computations up to one order of magnitude (as verified in our experiments).
At the beginning (Lines~\ref{alg:lb:start}--\ref{alg:lb:end}),  Algorithm~\ref{alg:lb} places each vertex $v$ in the bucket corresponding to
the value of $LB_2(v)$, and sets the flag $setLB(v)$ to true. Having $setLB(v)=true$ means that, for the vertex $v$,
the value of $deg^h_{G[V]}(v)$ is still not computed, but only the value of $LB_2(v)$ is known.
Next, it calls Algorithm~\ref{alg:partition}.

Algorithm~\ref{alg:partition} extracts a vertex $v$ from $B$ (Line~\ref{alg:partition:bucket}), it checks if $setLB(v)$ is equal to true: if yes,
it computes $deg^h_{G[V]}(v)$, moves $v$ to $B[deg^h_{G[V]}(v)]$, and sets $setLB(v)$ to false. Otherwise, it assigns $core(v)=k$,
finds all vertices $u$ in its $h$-neighborhood, and updates the value of $deg^h_{G[V]}(u)$ only for the neighbors $u$ for which $setLB(u)$ is false.
Moreover, if a vertex $u$ (whose $setLB(u)$ is false) is exactly at distance $h$ from a deleted vertex $v$, the algorithm just decreases the value of $u$'s
$h$-degree by 1 (Line~\ref{alg:partition:decrease}).
\begin{algorithm}[t!]
\caption{$h$\textsf{-LB}}
\label{alg:lb}
\begin{algorithmic}[1]
\REQUIRE graph $G=(V,E)$, distance threshold $h >1$
\ENSURE core index $core(v)$ for each vertex $v \in V$
\FORALL{$i=1,2,\ldots,|V|$}
\STATE $B[i]\gets \emptyset$
\ENDFOR
\FORALL{ $v \in V$\label{alg:lb:start}}
	\STATE  $LB_1(v)\leftarrow deg^{\lfloor\frac{h}{2}\rfloor}_{G[V]}(v)$
\ENDFOR
\FORALL{ $v \in V$\label{alg:lb:start1}}
	\FORALL{$u \in N_{G[V]}(v,\lceil\frac{h}{2}\rceil)$}
      \STATE $LB_2(v)\leftarrow \max\left(  LB_1(u), LB_1(v) \right) $
	\ENDFOR
    \STATE $setLB(v) \leftarrow true$
	\STATE $B[LB_2(v)]\leftarrow B[LB_2(v)] \cup \{v\}$\label{alg:lb:end}
\ENDFOR
\STATE $CoreDecomp(G,h,1,|V|,B,setLB)$ (Alg.~\ref{alg:partition})
\end{algorithmic}
\end{algorithm}
\begin{algorithm}[t!]
\caption{$CoreDecomp$}
\label{alg:partition}
\begin{algorithmic}[1]
\REQUIRE graph $G=(V,E)$, distance threshold $h >1$, $k_{min}, k_{max}\in \mathbb{N}^+$, bucket $B$, flags $setLB$
\ENSURE Core index  $\forall v \in V$ s.t. $k_{min}\leq core(v)\leq  k_{max}$
\FORALL{$k=k_{min}-1, k_{min}, \ldots,k_{max}$}
  \WHILE{$B[k] \neq \emptyset$}
	\STATE pick and remove a vertex $v$ from $B[k]$\label{alg:partition:bucket}
	  \IF{$setLB(v)$}
		\STATE compute $deg^h_{G[V]}(v)$
		\STATE $B[deg^h_{G[V]}(v)]\leftarrow B[deg^h_{G[V]}(v)] \cup \{v\}$
		\STATE $setLB(v) \leftarrow false$
	  \ELSE
		\IF{$k\geq k_{min}$}\label{alg:partition:ifkmin}
		  \STATE $core(v) \leftarrow k$		
		  \STATE $setLB(v) \leftarrow true$	\label{alg:partition:ifkmin:end}
		\ENDIF
		\FORALL{$u \in N_{G[V]}(v,h)$}\label{alg:partition:update}
		  \IF{not $setLB(u)$}
		    \IF{$d(u,v)_{G[V]}<h$}
		      \STATE compute $deg^h_{G[V\setminus\{v\}]}(u)$
		    \ELSE
		      \STATE $deg^h_{G[V\setminus\{v\}]}(u) = deg^h_{G[V\setminus\{v\}]}(u) -1 $\label{alg:partition:decrease}
		    \ENDIF
		    \STATE move u to $B[\max \left(deg^h_{G[V\setminus\{v\}]}(u), k\right)]$
		  \ENDIF
		\ENDFOR
		\STATE $V \leftarrow V\setminus\{v\}$
	  \ENDIF		
	\ENDWHILE
\ENDFOR
\end{algorithmic}
\vspace{-1mm}
\end{algorithm}

\spara{Correctness.}
Let $\tilde{V}$ and $\tilde{k}$ denote the current status of $V$ and $k$, respectively. At the beginning of every iteration of the outer for-loop (lines 3--8), it holds that $v \in B[i] $ if either $deg^h_{G[\tilde{V}]}(v) = i$ or $LB(v) = i$ and $setLB(v)=true$.
When the vertex $v$ is extracted, we check the status of the $setLB$ variable: Observation~\ref{th:lb2} ensures that, if we extract a vertex $v$ from $B[\tilde{k}]$  and  $setLB(v)$ is true, then $core(v)\geq \tilde{k}$.
Then, if $setLB(v)$ is true we compute the current value of $deg^h_{G[\tilde{V}]}(v)$ and we insert $v$ in $B[deg^h_{G[\tilde{V}]}(v)]$. On the other hand, if  $setLB(v)$ is false, we assign the core index as in Algorithm $h$\textsf{-BZ}.
In this latter case, we recompute only the $h$-degree of the $h$-neighbors $u$ of $v$ in $\tilde{V}\cup \{v\}$ such that $setLB(u)=false$ because, by Observation~\ref{th:lb2}, $core(u)\geq \tilde{k}$. Notice that, in the case $core(u)$ is exactly equal to $\tilde{k}$, $u$ is already in $B[\tilde{k}]$.

%
%
%
%
%
%
%
%
%
%
%
%

\spara{Computational complexity.}
The time complexity is asymptotically the same of the $h$\textsf{-BZ} algorithm; however, as we will show in Section \ref{sec:experiments}, $h$\textsf{-LB} is typically much faster thanks to the reduced number of $h$-degree re-computations.
\subsection{Lower and upper bound algorithm}
\label{sec:topdown_algo}

The vertices incurring the largest cost for the computation of the $(k,h)$-core decomposition, are the ones forming the cores with large values of $k$: those are vertices with extremely large $h$-neighborhood, which needs to be updated a large number of times (i.e., each time a neighbor in the lower cores is removed), and for which any update requires a large $h$-bounded BFS traversals. Although the lower-bound mechanism, introduced in the $h$\textsf{-LB} algorithm, partially alleviates this overhead, these vertices still are responsible for the largest chunk of the computation.
The algorithm we introduce next exploits an upper bound on the core index of each vertex to partition the computation in a set of sub-computations which are totally independent from each other. This way it can adopt a \emph{top-down}\footnote{\scriptsize Cheng et al.~\cite{cheng2011efficient} devised an external memory algorithm with a similar top-down approach.
However, their method has been designed for the classic core decomposition (i.e., $h=1$), hence not immediately applicable to our problem.}
 explorations of the cores (from larger to smaller values of $k$):  by discovering and peeling  the vertices with high core index at earlier stages of the algorithm, many costly $h$-bounded BFS traversals are saved.

We first present the overall logic of the algorithm, named $h$\textsf{-LB+UB},
then we introduce the specific upper bound used (\S\ref{subsubsec:computingUB}). Later we show how, thanks to the partitioning of the computation, we can have a tighter lower bound (\S\ref{subsubsec:improveLB}). Finally we discuss how to parallelize the computations of $h$-degrees, exploiting multi-threading (\S\ref{subsubsec:para}).

For a given $h> 1$, suppose we have an upper bound on the core index of each vertex. i.e., a function $UB: V \rightarrow \mathbb{N}$.
Let $U = \{ub_1, \ldots, ub_\ell\}$ be the ordered codomain of $UB$. For any $i \leq ub_\ell$ let
$V[i] = \{ v \in V | UB(v) \geq i\}$, and $G[V[i]]$ be the subgraph induced by $V[i]$ . The following holds.

\begin{observation}\label{obs}
All $(k,h)$-cores with $k \geq i$ are contained in $V[i]$ and thus can be computed from $G[V[i]]$.
\end{observation}

Observation~\ref{obs} holds by virtue of upper bound on the core index.
Based on this observation we can split the computation of the distance-generalized core decomposition in a set of sub-computations which are totally independent from each other. In particular, we could consider any partition in contiguous intervals of $[lb_0,ub_\ell]$ where $lb_0 = min_{v \in V} LB_2(v)$ is the minimum lower bound  in the interval, and $ub_\ell$ is the maximum upper bound in the interval. Then for each of these intervals $[i,j]$ we will search for the $(k,h)$-cores with $i \leq k \leq j$ in the subgraph $G[V[i]]$. It is important to note that in the higher intervals, i.e., where we seek for the $(k,h)$-cores with higher $k$, we deal with a smaller set of vertices (as $V[i] \supseteq V[j]$ for $i < j$).

Among all possible ways of partitioning the computation, algorithm $h$\textsf{-LB+UB} creates intervals covering $S$ contiguous values of upper bounds (i.e., elements of $U$), where $S \in \mathbb{N}^+$ is an input parameter. As already anticipated, the intervals are visited in a top-down fashion.
\begin{myexample}
Suppose that in a graph we have upper bounds $U = \{5,10,15,20,25,30\}$, $lb_0 = 3$  and $S = 2$. Algorithm $h$\textsf{-LB+UB} would partition the computation as follows: $\langle [30,21],[11,20],[3,10]\rangle$. Instead for $S = 1$ it would be $\langle [30,26],[21,25],[16,20],[11,15],[6,10],[3,5]\rangle$.
\end{myexample}

Inside each partition the vertices are processed the same way they were in
previous $h$\textsf{-LB} approach, i.e., by means of Algorithm~\ref{alg:partition}.
Notice that for the vertices with $core()<k_{min}$, the condition in Line~\ref{alg:partition:ifkmin} of Algorithms~\ref{alg:partition} is false,
so their core indices are not assigned (their core indices will be assigned in a subsequent partition).

The whole process is
repeated for all the partitions in order (Lines~\ref{alg:topdown:parpart:start}--\ref{alg:topdown:parpart:end}, Algorithm~\ref{alg:topdown}).
Before processing the vertices, in Line~\ref{alg:topdown:clean}, we run Algorithm~\ref{alg:clean} (i.e., $ImproveLB$)
which removes unimportant vertices from $V[k_{min}]$ and returns a new lower bound for each vertex in $V[k_{min}]$.

\begin{algorithm}[t!]
\caption{$h$\textsf{-LB+UB}}
\label{alg:topdown}
\begin{algorithmic}[1]
\REQUIRE $G=(V,E)$, threshold $h >1$, partition size $S \in \mathbb{N}^+$
\ENSURE core index $core(v)$ for each vertex $v \in V$
\FORALL{$i=1,2,\ldots,|V|$}
\STATE $B[i]\gets \emptyset$
\ENDFOR
\FORALL{$v \in V$}\label{alg:topdown:h-degree:start}
  \STATE compute $deg^h_{G[V]}(v)$
  \STATE compute $LB_2(v)$
    \STATE $LB_3(v)\gets 0$
\ENDFOR\label{alg:topdown:h-degree:end}
\STATE $UpperBound(G,h)$ $\;\;\;$ (Algorithm~\ref{alg:ub_algo})
\STATE $U\gets \{UB(v) : v \in V\}$
\STATE $U \gets  U\cup min_{v\in V}(LB_2(v)-1)$
\STATE sort $U$ in \textbf{descending} order
\FORALL{$(k_{min},k_{max})\in\{(U[0],U[S]+1),(U[S],U[2S]+1), \ldots, (U[\lfloor\frac{|U|-1}{S}\rfloor S -S],U[|U|-1])\}$}\label{alg:topdown:parpart:start}


  \STATE $V[k_{min}] \gets \{v \in V | UB(v)\geq k_{min}\}$~\label{alg:topdown:partition}
  \STATE $V[k_{min}],LB_3^* \gets ImproveLB(V[k_{min}],h,k_{min})$ \label{alg:topdown:clean};
   \STATE $LB_3(v) \gets \max\{LB_3(v),LB_3^*(v)\} \forall v \in V[k_{min}] $
   \FORALL{$v \in V[k_{min}]$}\label{alg:topdown:lbrefine:start}
      \STATE Add $v$ to $B[\max(core(v),LB_3(v),k_{min}-1)]$
      \STATE $setLB(v) \leftarrow true$
   \ENDFOR\label{alg:topdown:lbrefine:end}
   \STATE $CoreDecomp(G[V[k_{min}]],h,k_{min},k_{max},B,setLB)$
\ENDFOR
\label{alg:topdown:parpart:end}\label{alg:topdown:peeling:end}
\end{algorithmic}
\end{algorithm}
\begin{algorithm}[t!]
\caption{$UpperBound$}
\label{alg:ub_algo}
\begin{algorithmic}[1]
\REQUIRE $G=(V,E)$, distance threshold $h >1$
\ENSURE Upper bound $UB(v)$ for each vertex $v \in V$
\FORALL{$i=1,2,\ldots,|V|$}
\STATE $B[i]\gets \emptyset$
\ENDFOR
\FORALL{$v \in V$}
\STATE compute $deg^h_{G[V]}(v)$
\STATE  $UBdeg^h_{G[V]}(v) \gets deg^h_{G[V]}(v)$
\STATE $B[UBdeg^h_{G[V]}(v)]\leftarrow B[UBdeg^h_{G[V]}(v)] \cup \{v\}$
\ENDFOR
\FORALL{$k=1,2,\ldots,|V|$}
\WHILE{$B[k] \neq \emptyset$}
\STATE pick and remove a vertex $v$ from $B[k]$
\FORALL{$u \in N_{G[V]}(v,h)$}\label{alg:ub_algo:neighbors}
\STATE $UBdeg^h_{G[V]}(u) = UBdeg^h_{G[V]}(u) - 1$\label{alg:ub_algo:decrease}
\STATE move u to $B[\max \left(UBdeg^h_{G[V]}(u), k\right)]$
\ENDFOR
\ENDWHILE
\ENDFOR
\end{algorithmic}
\end{algorithm}

\spara{Correctness.}
All vertices with $UB< k_{min}$ do not belong to $V[k_{min}]$. Assuming the correctness of $UB$ computation,
those vertices not in $V[k_{min}]$ must have core indexes smaller than $k_{min}$. Now, the correctness of
Algorithm~\ref{alg:topdown} directly follows from that of $h$\textsf{-LB}: indeed Algorithm~\ref{alg:topdown} is correct for any lower bound on the core indexes of the vertices. In particular, for each vertex $v\in V[k_{min}]$, Algorithm~\ref{alg:topdown} correctly assigns the core index
of those vertices $v$ s.t. $k_{min}\leq core(v) \leq k_{max}$ by Observation~\ref{obs}.

\spara{Computational complexity.}
The time complexity is asymptotically the same of the $h$\textsf{-BZ} algorithm; however, as we shall demonstrate
in Section \ref{sec:experiments}, $h$\textsf{-LB+UB} is typically much faster thanks to the reduced number of $h$-degree
re-computations for the vertices belonging to the inner most cores. $h$\textsf{-LB+UB} allows to speed-up the computation up to one order of magnitude in graphs with more than a million of vertices.
We shall later show that one can further improve the efficiency
by parallelizing some blocks of our algorithms.

\subsection{Computing the upper bound}\label{subsubsec:computingUB}
We next present our method to efficiently compute a good upper bound. In Section 1 we have shown that performing a classic core decomposition of the power graph $G^h$ does not provide the correct $(k,h)$-core decomposition. However, it turns out that the core index that we compute this way for each vertex, is an upper bound of its $(k,h)$-core index. This is the key idea at the basis of  Algorithm~\ref{alg:ub_algo}.

One challenge is that materializing the power graph $G^h$ might result in a graph too large to fit in memory.
For this reason we avoid keeping the $h$-neighborhoods in memory. This forces us to recompute, each time we remove a vertex, its neighborhood
at Line~\ref{alg:ub_algo:neighbors}: these are the vertices whose approximated $h$-degree  ($UBdeg^h$) is decreased by 1 at Line~\ref{alg:ub_algo:decrease}.
However, as we know, when we remove a vertex, the real $h$-degree of its $h$-neighbors can potentially decrease of more than 1: this is why what we get is an upper bound and not the correct core index.

\subsection{Improving the lower bound}\label{subsubsec:improveLB}
Next we discuss Algorithm~\ref{alg:clean}, which is invoked at Line~\ref{alg:topdown:clean} of the $h$\textsf{-LB+UB} method.
As said earlier, it computes a new (tighter) lower bound $LB_3$ for each vertex in $V[k]$.
The correctness of $LB_3$ is ensured by Property~\ref{prop:mindegree}.
\begin{myproperty}\label{prop:mindegree}
Given a graph $G=(V,E)$ and a distance threshold $h > 1$, it holds that for any
$V'\subseteq V$ and any $u \in V'$
$$\min(deg^h_{G[V']}(v)| v\in V') \leq core(u),$$
where $core(u)$ is the core index of $u$ in the original graph  $G$.
\end{myproperty}
%
%
%

Intuitively, Property~\ref{prop:mindegree} holds because
every vertex $u \in V'$ must have $h$-degree at least $d^*=\min(deg^h_{G[V']}(v)| v\in V')$ in the subgraph $G[V']$.
Then, by definition of $(k,h$)-core, every $u \in V'$ must be in $(d^*, h)$-core of $G[V']$.
Since $G[V']$ is a subgraph of $G$, it follows that $core(u)\geq core_{G[V']}(u)$.
In our case, $\min(deg^h_{G[V[k]]}())$ is a lower bound on the core indexes for each vertex in $V[k]$.
By considering the maximum between this lower bound and $LB_2$, we achieve a tighter lower
bound $LB_3$ for every vertex in $V[k]$, which is exploited in the subsequent processing of
the vertices in $V[k]$ (Lines \ref{alg:topdown:lbrefine:start}--\ref{alg:topdown:lbrefine:end}, Algorithm~\ref{alg:topdown}). Since $LB_3$ is often tighter than $LB_2$, $setLB$ remains true for a larger number of iterations, thus allowing to save many $h$-degree re-computations

Algorithm~\ref{alg:clean} also efficiently and effectively ``cleans'' the set $V[k]$, often emptying it (i.e., when the partition does not contain any core). In particular, vertices with core index definitely smaller than $k_{min}$ are removed in lines \ref{alg:clean:delete:start}--\ref{alg:clean:delete:end}:
More in details, Algorithm~\ref{alg:clean} iteratively deletes vertices with $deg^h_{G[V_k]}()<k_{min}$, following the same power-graph idea used in Algorithm~\ref{alg:ub_algo}. For every vertex deletion, it only decrease by 1 the $h$-degree of their neighbors obtaining an upper bound on the effective $h$-degree: it is straightforward that, if a vertex has the upper bound of the $h$-degree smaller than $k_{min}$, it does not belong to the current partition.
\setlength{\textfloatsep}{14pt}
\begin{algorithm}
\caption{$ImproveLB$}
\label{alg:clean}
\begin{algorithmic}[1]
\REQUIRE set of vertices $V[k]$,  $h >1$,  $k \in \mathbb{N}^+$
\ENSURE $V'[k]\subseteq V[k]$,  lower bound $LB_3(v) \;  \forall v \in V$
\STATE $D\gets \emptyset$
\FORALL{$v \in V[k]$}\label{alg:clean:init}
\STATE compute $deg^h_{G[V[k]]}(v)$
\IF{$deg^h_{G[V[k]]}(v)<k$}
\STATE $D\gets D \cup v$
\ENDIF
\ENDFOR
\STATE $min(deg^h_{G[V[k]]})\gets \min\{v\in V[k]|deg^h_{G[V[k]]}(v)\}$
\FORALL{$v \in V[k]$}
\STATE $LB_3(v) \gets \max\{LB_2(v),min(deg^h_{G[V[k]]})\}$\label{alg:clean:ub}
\ENDFOR
\WHILE{$D\neq \emptyset$}\label{alg:clean:delete:start}
\STATE pick and remove a vertex $v$ from $D$
\STATE {$V[k] \gets V[k]\setminus \{v\}$}
\FORALL{$u \in N_{G[V[k]]}(v,h)$}
\STATE  $deg^h_{G[V[k]]}(u) \gets deg^h_{G[V[k]]}(u) - 1$
\IF{$deg^h_{G[V[k]]}(u)<k $}
\STATE $D \gets D \cup \{u\}$ \label{alg:clean:delete:end}
\ENDIF
\ENDFOR
\ENDWHILE
\RETURN $V[k]$ and $LB_3(v) \;  \forall v \in V$

\end{algorithmic}

\end{algorithm}

Next example explains the benefits of $h$\textsf{-LB+UB} (Algorithm~\ref{alg:topdown}) over
$h$\textsf{-LB} (Algorithm \ref{alg:lb}) and demonstrates the effectiveness of the
``cleaning'' procedure in Algorithm \ref{alg:clean}.
\begin{myexample}
Consider again the graph in Figure~\ref{fig:firstexemple},  and assume $h=2$.
$h$\textsf{-LB} starts by computing the $LB_2$ bound for all vertices, and placing
the vertices in the respective buckets, i.e., $B[2]=\{v_1\}$, and $B[5]=\{v_2,v_3,v_4,v_5,v_6, v_7,v_8,v_9,v_{10},v_{11},v_{12}, v_{13}\}$.
The $SetLB$ flag for each vertex is set as true.
We pick $v_1$ from $B[2]$, compute its $h$-degree, and move it to $B[4]$, and we set $setLB(v_1)=false$.
Then, we select again $v_1$ from $B[4]$ and we find that its $2$-degree is same as its $LB_2$ bound (since $setLB(v_1)=false$);
therefore, we assign its core index as $4$, and remove $v_1$ from the set of active vertices.
All the vertices in the $2$-hop neighborhood of $v_1$ have their $SetLB$ flags as true, and
it is not necessary to re-compute the 2-degree of them (because they are in $B[5]$, while we are currently processing $B[4]$).
Now, $B[5]$ has 12 vertices $v_2, v_3, \ldots, v_{13}$, and all of their $SetLB$ flags are true, thus one can process
them in any arbitrary order. Let us assume that we process them in descending order of their ids, i.e., we process
$v_{13}$ at first, and $v_2$ at the end. This will result in computation of their $h$-degrees, and we move
all but $v_3$ and $v_2$ to $B[6]$, while $v_3$ and $v_2$ will remain in $B[5]$. All the $setLB$ flags are
now false. Next, we select again $v_3$ from $B[5]$ and, since $setLB(v_3)=false$, we assign its core index as $5$, remove $v_3$ from the set of active vertices
and recompute the $h$-degree for their 2-neighbors. We skip the rest of the execution.

\begin{samepage}
We now show the execution of the $h$\textsf{-LB+UB} Algorithm. It starts computing the values of $UB$ using Algorithm~\ref{alg:ub_algo}	i.e $UB(v_1)=4$ and, for $2\leq i \leq 13$, $UB(v_i)=6$. Then, we create the first partition $V_6$ with $k_{min}=k_{max}=6$ with all the vertices $v_i$ s.t. $2\leq i \leq 13$.
Next, we run Algorithm~\ref{alg:clean} on the subgraph induced by $V_6$ and it removes the vertices $v_2$ and $v_3$.
This is because the 2-degree of $v_2$ and $v_3$ in the subgraph induced by $V_6$ is 5, which is smaller than the current
$k_{min}$=6.
Then, we re-compute the $2$-degree in the new subgraph, and run Algorithm~\ref{alg:partition} on the subgraph induced by the vertices in
$V_6 \setminus \{v_2, v_3\}$, thus we assign the core index 6 to them. Next, we process the subgraph induced by $V_5 \cup V_6$, assigning to $v_2$ and $v_3$'s
core index as 5. Since the core index of the vertices in $V_6$ are already computed, the peeling of $v_2$ and $v_3$ does not imply the re-computation of the 2-degree
of the vertices in $V_6$.
Notice that, while running $h$-LB, after the deletion of $v_3$, we need to recompute the $h$-degree of their neighbors.
This demonstrates the higher efficiency of $h$\textsf{-LB+UB} over $h$-LB. For space economy, we skip the rest of the execution.\end{samepage}
\end{myexample}

\subsection{Multi-threading}\label{subsubsec:para}
We have seen how Algorithm~\ref{alg:topdown} divides the computation in a number of sub-tasks totally independent from each other. As such one could parallelize their execution. However, while their independent execution is indeed possible, ``lower'' intervals could benefit from the knowledge coming from already completed ``higher'' intervals, i.e., the vertices with high core index that have been already processed and for which we do not have to keep updating the $h$-degree. By losing this knowledge we also lose the opportunity of having tighter $LB_3$ bounds. Therefore we are facing a trade-off: on the one hand we have the benefit of the parallel execution of different fractions of the computation, on the other hand we have the benefits of the top-down execution which are partially lost when parallelizing.

Another option of parallelizing certain blocks of our algorithms, is trivially to give different $h$-BFS traversals to different processors.
Let $C$ be the number of available processors. In the initial computation of $h$-degree
(Lines~\ref{alg:topdown:h-degree:start}--\ref{alg:topdown:h-degree:end}, Algorithm~\ref{alg:topdown}), we create a thread for each processor, we assign $\frac{|V|}{C}$
vertices to each thread, and we perform, in parallel, an $h$-BFS from each vertex.
Each vertex is dynamically assigned to some thread in order to balance the load among the processors.
We parallelize in the same way the initial $h$-degree computation in Algorithm~\ref{alg:clean}, Line~\ref{alg:clean:init}.
Then, we parallelize the update of $h$-degrees of the $h$-neighbors
(Algorithm~\ref{alg:partition}, Line~\ref{alg:partition:update}), by assigning dynamically
to each thread, $\frac{|N_{G[V]}(v,h)|}{C} $ neighbors.
To avoid race conditions, we consider the update of bucket $B$ as an atomic operation.

We empirically tried both parallelization options and found the second one to perform better: this is the one implemented in our algorithm.

\section{Applications}
\label{sec:applications}
In this section we show that the distance-generalized core decomposition preserves several nice features of the classic core decomposition,
and can be used to speed-up or approximate distance-generalized notions of dense structures.

First (\S\ref{subsec:chromatic}), we demonstrate,  to the distance-generalized case, a connection existing between the maximum core index of a graph and its \emph{chromatic number}. Then  (\S\ref{subsec:club}), we tackle the \emph{maximum $h$-club} problem and we show how to exploit the distance-generalized core decomposition to speed-up these existing methods for this hard problem. Finally (\S\ref{subsec:densest}), we introduce the novel problem of \emph{distance-generalized densest subgraph} and prove that by using distance-generalized core decomposition, one can obtain an efficient algorithm with approximation guarantees.

 In Appendix \ref{subsec:cocktail}, we show an additional application: the distance-based generalization of \emph{cocktail-party problem}~\cite{sozio2010}.


\subsection{Distance-$h$ chromatic number}
\label{subsec:chromatic}
The distance-$h$ chromatic number is a generalization of the classical notion of chromatic number,
and it was introduced in the eighties by McCormick \cite{McCormick1983}.
\begin{mydefinition}[Distance-$h$ chromatic number ]
A \emph{distance-$h$ coloring} of a graph $G=(V,E)$ is a partition of $V$
into classes (colors) so that any two vertices of the same color are more than $h$ hops apart.
The {\em distance-$h$ chromatic number}, $\chi_h(G)$ is the minimum number of colors required
so that any two vertices having the same color are more than $h$ hops apart.
\end{mydefinition}

This is useful in scheduling, register allocation,
finding the optimal seating plan in an event, e.g., (a) find the minimum number of registers required for concurrently storing
variables that are accessed within a span of $h$ subroutine calls during the execution of a program, and (b) find the minimum
number of sessions required in a courtroom such that no two persons, who are socially connected within $h$-distance of each other, are present
in the same session, etc. This is also related to the
{\em distance-$h$ independent set} problem \cite{CN84}: given a graph $G=(V,E)$, $I\subseteq V$ is a
distance-$h$ independent set if for every distinct pair $i,j \in I$, $d(i,j)\ge h+1$.

McCormick \cite{McCormick1983} proved
that finding the distance-$h$ chromatic number is \NP-hard, for any fixed $h\ge 2$.
We next generalize a relation existing between the classic notion of chromatic number and the classic core decomposition \cite{SZEKERES19681,MatulaB83} to their distance-generalized versions.

Let us denote by $h$-degeneracy $\hat{C}_h(G)$ of a graph $G$, the largest value $k$ such that it has a non-empty $(k,h)$-core.
We next prove that $\hat{C}_h(G)$ provides an upper bound on the distance-$h$ chromatic number, $\chi_h(G)$.
\begin{mytheorem} \label{th:chromatic}
$
\chi_h(G) \le 1+\hat{C}_h(G).
$
\end{mytheorem}
%
%
%

\subsection{Maximum $h$-club}\label{subsec:club}
In \S\ref{sec:related} we introduced some background information regarding the problem of computing a maximum $h$-club.
We next formalize some of these concepts, and provide a characterization theorem connecting the maximum size of an $h$-clique and an $h$-club
with the distance-$h$ chromatic number and the maximum $k$ of the $(k,h)$-core decomposition.
\begin{mydefinition}[$h$-clique]
Given a graph $G = (V,E)$ and a distance threshold $h >1$, an $h$-clique is a subset of vertices  $S \subseteq V$ such that $d_G(u,v)\le h, \; \forall u,v \in S $.
An $h$-clique is said to be \emph{maximum} if there is no larger $h$-clique. We denote $\tilde{w}_h(G)$ the cardinality of a maximum $h$-clique.
\end{mydefinition}
\begin{mydefinition}[$h$-club]
Given a graph $G = (V,E)$ and a distance threshold $h > 1$, an $h$-club is a subset of vertices  $S \subseteq V$  such that $d_{G[S]}(u,v)\le h, \, \forall u,v \in S$.
An $h$-club is said to be \emph{maximum} it there is no larger $h$-club. We denote $\hat{w}_h(G)$  the cardinality of a maximum $h$-club.
\end{mydefinition}
Clearly, every $h$-club is contained in some $h$-clique. Let $w(G)$
denote the cardinality of a maximum clique in $G$. Then, the following inequality holds:
$
w(G) \le \hat{w}_h(G) \le \tilde{w}_h(G).
$
Moreover, in any distance-$h$ coloring of a graph $G$,
an $h$-clique can intersect any color class in at most one vertex.
Combining this observation with Theorem~\ref{th:chromatic}, we obtain the following result.
\begin{mytheorem} \label{th:characterization}
$$
w(G) \le \hat{w}_h(G) \le \tilde{w}_h(G) \le \chi_h(G) \le 1+\hat{C}_h(G).
$$
\end{mytheorem}
As already discussed in \S\ref{sec:related}, the problems of finding $\hat{w}_h(G)$ and $\tilde{w}_h(G)$ are \NP-hard, and the \emph{maximum $h$-club problem} is the hardest of the two due to the fact that $h$-clubs are not closed under set inclusion.
We next show how to use the proposed $(k,h)$-core decomposition
 to speed-up the methods for the maximum $h$-club problem.
We note that various exact and heuristic methods
were developed in the literature for identifying $h$-clubs \cite{AC12,Balasundaram2005,CHLS13,BLP00,SKMN12,SB13,VPP15,VEREMYEV2012316,MB15}.
Our proposal is orthogonal and complementary to this literature, in fact, it can be used in conjunction with any of these algorithms. We exploit the following observation.
\begin{mytheorem} \label{th:core_club}
Every $h$-club of size $k+1$ must be included in the $(k,h)$-core, $\mathcal{C}(k,h)$, of a graph $G$.
\end{mytheorem}
%
%

%
Following this observation, we can use the $(k,h)$-core decomposition as a wrapper around any black-box algorithm from the literature which takes as input a graph $G$ and a parameter $h > 1$, and return the maximum $h$-club in $G$. Denote such black-box algorithm $\mathcal{A}(G,h)$. Our proposed algorithm simply starts by computing the  $(k,h)$-core decomposition of $G$ (using one of the algorithms that we introduced earlier in \S~\ref{sec:algorithms}). Let $C_{k^*}$ be the core of maximum index, our method first invokes $\mathcal{A}(G[ C_{k^*}],h)$, which is much faster and less memory consuming than $\mathcal{A}(G,h)$, since $G[ C_{k^*}]$ is smaller than $G$. If an $h$-club of size $S> k^*$ is found, then this is the maximum $h$-club (following Theorem~\ref{th:core_club}) and the algorithm terminates. Otherwise, we search in the lower core, i.e., by invoking $\mathcal{A}(G[ C_{min\{S,k^*-1\}}],h)$, and so on, until we find an $h$-club of size larger than the index of the current core.
The pseudocode of our approach is given in Algorithm~\ref{alg:h-club} (Appendix \ref{appe:pc}).

\subsection{Distance-generalized densest subgraph}
\label{subsec:densest}
As discussed in  \S\ref{sec:related} the most well-studied notion of a dense subgraph is the one maximizing the average degree, which is known as \emph{densest subgraph}~\cite{Goldberg84}. We next generalize this notion by considering the average $h$-degree.
\begin{problem}[Distance-$h$ densest subgraph]\label{prob:densestsubgraph}
Given a graph $G=(V,E)$ and a distance threshold $h \in \mathbb{N}^+$,  find a subset $S^* \subseteq V$ with the maximum average $h$-degree.
$$
 S^*=\underset{S\subseteq V}{\argmax}\frac{\sum_{v\in S} deg^h_{G[S]}(v)}{|S|}
$$
\end{problem}
It is easy to see that for $h = 1$, Problem~\ref{prob:densestsubgraph} corresponds to the traditional densest-subgraph problem in simple graphs~\cite{Goldberg84}.
Such a problem is solvable in polynomial time, but the time complexity of the exact algorithm  is $\Omega(|V| \times |E|)$~\cite{Goldberg84}, thus unaffordable for large graphs.
Hence, we cannot hope scalable exact solutions to exist for Problem~\ref{prob:densestsubgraph} with $h > 1$ either.
Analogously to what has been done for the densest-subgraph problem ($h=1$)~\cite{AITT00,Char00}, among all
cores obtained via the $(k,h)$-core decomposition, we use the core which exhibits the maximum average $h$-degree as an approximation of the distance-$h$ densest subgraph.
The quality of the approximation is guaranteed by the following theorem.
\begin{mytheorem}\label{th:densest}
We denote by $G[S^*]$ the distance-$h$ densest subgraph, and its average $h$-degree as $f_h(S^*)$. We denote by $C_{k}$
the core with the maximum average $h$-degree , among all other cores obtained via the $(k,h)$-core decomposition.
Then, the core $C_{k}$ provides  $(\sqrt{f_h(S^*)+0.25}-0.5)$-approximation to the distance-$h$ densest subgraph problem, i.e.,
$$
f_h(C_{k})\ge (\sqrt{f_h(S^*)+0.25}-0.5)
$$
\end{mytheorem}

\section{Experimental Assessment}
\label{sec:experiments}
\begin{table}[t!]
	\centering
		\caption{Characteristics of datasets used.}
\vspace{-3mm}
	\label{tab:graphs}
	\small
	\begin{tabular}{r|c|c|c|c|c|}
	\multicolumn{1}{c}{}	 & \multicolumn{1}{c}{$|V|$} & \multicolumn{1}{c}{$|E|$} & \multicolumn{1}{c}{avg deg} & \multicolumn{1}{c}{max deg} & \multicolumn{1}{c}{diam} \\
		\cline{2-6}
		\texttt{coli} & 328 & 456  &2.78&100&14\\
		\texttt{cele} & 346 & 1,493 &8.63&186  & 7\\
				\texttt{jazz} & 198 & 2,742  &27.70&100 & 6\\
		\texttt{FBco} & 4,039 & 88,234  &43.69&1,045&8\\
		\texttt{caHe} & 11,204 & 117,619  &19.74&491&13\\
		\texttt{caAs} & 17,903 & 196,972  &21.10&504&14\\
						\texttt{doub} &154,908&327,162&4.22&287&9\\
				\texttt{amzn} & 334,863 &925,872  &3.38&549 & 44 \\
						\texttt{rnPA} & 1,090,920 & 1,541,898  &2.83&9&786 \\
		\texttt{rnTX} & 1,393,383 &1,921,660   &2.76&12&1,054\\
									\texttt{sytb} &495,957  &1,936,748   &3.91&25,409&21\\
		\texttt{hyves} & 1,402,673 &2,777,419 & 3.96  &31,883 &10\\
		\texttt{lj} &4,847,571&68,993,773 &14.23&14,815&16\\
		\cline{2-6}
	\end{tabular}
\vspace{-2mm}
\end{table}

We use thirteen real-world, publicly-available graphs whose characteristics are summarized in Table~\ref{tab:graphs}.
All graphs are undirected and unweighted:
\texttt{coli1}\footnote{\url{http://lasagne-unifi.sourceforge.net/}\label{footnote:lasagne}} (\texttt{coli} for short),
\texttt{celegans\_metab}\textsuperscript{\ref{footnote:lasagne}}  (\texttt{cele}) are biological networks;
\texttt{jazz}\footnote{\url{http://konect.uni-koblenz.de/}\label{footnote:konect}} is a collaboration network among jazz musicians;
\texttt{ca-HepPh}\textsuperscript{\ref{footnote:snap}} (\texttt{caHe}),
\texttt{ca-AstroPh}\textsuperscript{\ref{footnote:snap}} (\texttt{caAs}) are collaboration networks among scientists;
\texttt{facebook-comb}\footnote{\url{http://snap.stanford.edu/}\label{footnote:snap}} (\texttt{FBco}),
\texttt{douban}\textsuperscript{\ref{footnote:konect}} (\texttt{doub}),
\texttt{soc-youtube}\footnote{\url{http://networkrepository.com/}}  (\texttt{sytb}),
 \texttt{soc-livejournal}\textsuperscript{\ref{footnote:snap}} (\texttt{lj})
and \texttt{hyves}\textsuperscript{\ref{footnote:konect}} are social graphs;
\texttt{roadNet-PA}\textsuperscript{\ref{footnote:snap}} (\texttt{rnPA}),
\texttt{roadNet-TX}\textsuperscript{\ref{footnote:snap}} (\texttt{rnTX}) are road networks and
\texttt{com-amazon}\textsuperscript{\ref{footnote:snap}} (\texttt{amzn}) is a co-purchasing network.

\begin{table}[t!]
	\centering
	\caption{\small Maximum core index / number of distinct cores.}
	\label{tab:num_core}
\vspace{-3mm}
	\small
		\setlength\tabcolsep{1.9pt}
	\begin{tabular}{r|c|c|c|c|c|}
	\multicolumn{1}{c}{}	& \multicolumn{1}{c}{$h=1$}  & \multicolumn{1}{c}{$h=2$}  & \multicolumn{1}{c}{$h=3$}  & \multicolumn{1}{c}{$h=4$}  & \multicolumn{1}{c}{$h=5$}  \\
		\cline{2-6}
		\texttt{coli} & 3 / 3 & 72 / 20 & 85 /40 &139 / 32 &198 / 26\\
\texttt{cele} & 10 / 10 & 186 / 52 & 291 / 25 & 336 / 6 & 342 / 3 \\
				\texttt{jazz} &29 / 21 & 109 / 27 & 174 / 12 & 191 / 6 &196 / 2 \\
	\texttt{FBco} &115 / 96 &1045 / 43 & 1829 / 15 &3228 / 10 & 3777 / 5\\
	\texttt{caHe}  &238 / 65 &654 / 589 &2267 / 1678 & 4392 / 2121 & 7225 /1237 \\
		\texttt{caAs}  &56 / 53 &680 / 675 &4305 / 3339 &10252 / 2757 &14403 / 1185 \\
		\cline{2-6}
	\end{tabular}
\vspace{2mm}
\end{table}

All algorithms are implemented in \texttt{C++} using NetworKit framework\footnote{\url{http://networkit.iti.kit.edu/}}.
The experiments are conducted on a server with 52 cores (Intel 2.4 GHz CPU) with 128 GB RAM.

In all the experiments we use values of $h \in [2,5]$: \emph{larger values are not so interesting} because as $h$ approaches the diameter of the network, all the vertices become reachable from all the vertices, and only few cores with very high $k$ would end up containing all the vertices.
As an example, consider that the average distance between two people in Facebook is 4.74 \cite{BackstromBRUV12}.

The goals of our experimentation are: to characterize the $(k,h)$-cores (\S\ref{subsec:exp1}) for different values of $h$ in terms of number of cores and distribution of core indexes; to compare efficiency of different algorithms (\S\ref{subsec:exp2}); to study the effectiveness of the proposed lower and upper bounds (\S\ref{subsec:expbounds}); to assess scalability to larger graphs (\S\ref{subsec:expscalability}); and finally to showcase effectiveness in applications (\S\ref{subsec:exp3}-\ref{subsec:landmarks}).

\subsection{Characterization of the $(k,h)$-cores}
\label{subsec:exp1}
Table~\ref{tab:num_core} reports the number of $(k,h)$-cores for different values of $h$.
In particular, the left number is the maximum core index, while the right number counts how many of the cores are distinct.  We observe that by increasing the values of $h$ from 1 to $2-3$, the number of distinct cores grows substantially, capturing more structural differences among the vertices, and providing a finer-grained analysis.
On the other hand, for $h \geq 4$, while the maximum core index keeps growing, more and more vertices end up belonging to the same core as the network become more connected within $h$ steps. This is more evident for networks with smaller diameter.

\begin{figure}[h!t!]
\begin{tabular}{cc}
\texttt{caAs} & \texttt{FBco} \\
\hspace{-4mm}\includegraphics[width=.245\textwidth]{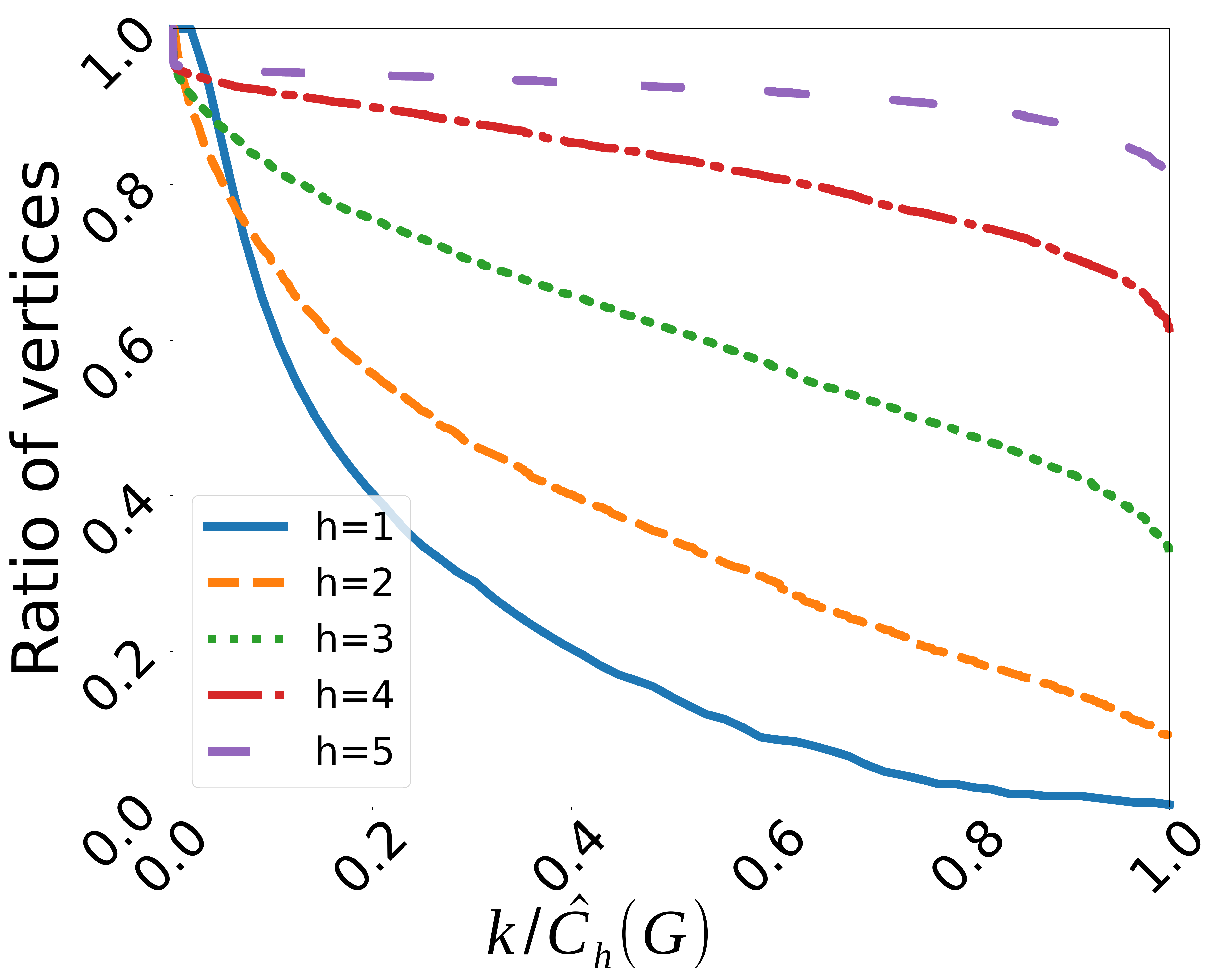} &
\hspace{-4mm}\includegraphics[width=.245\textwidth]{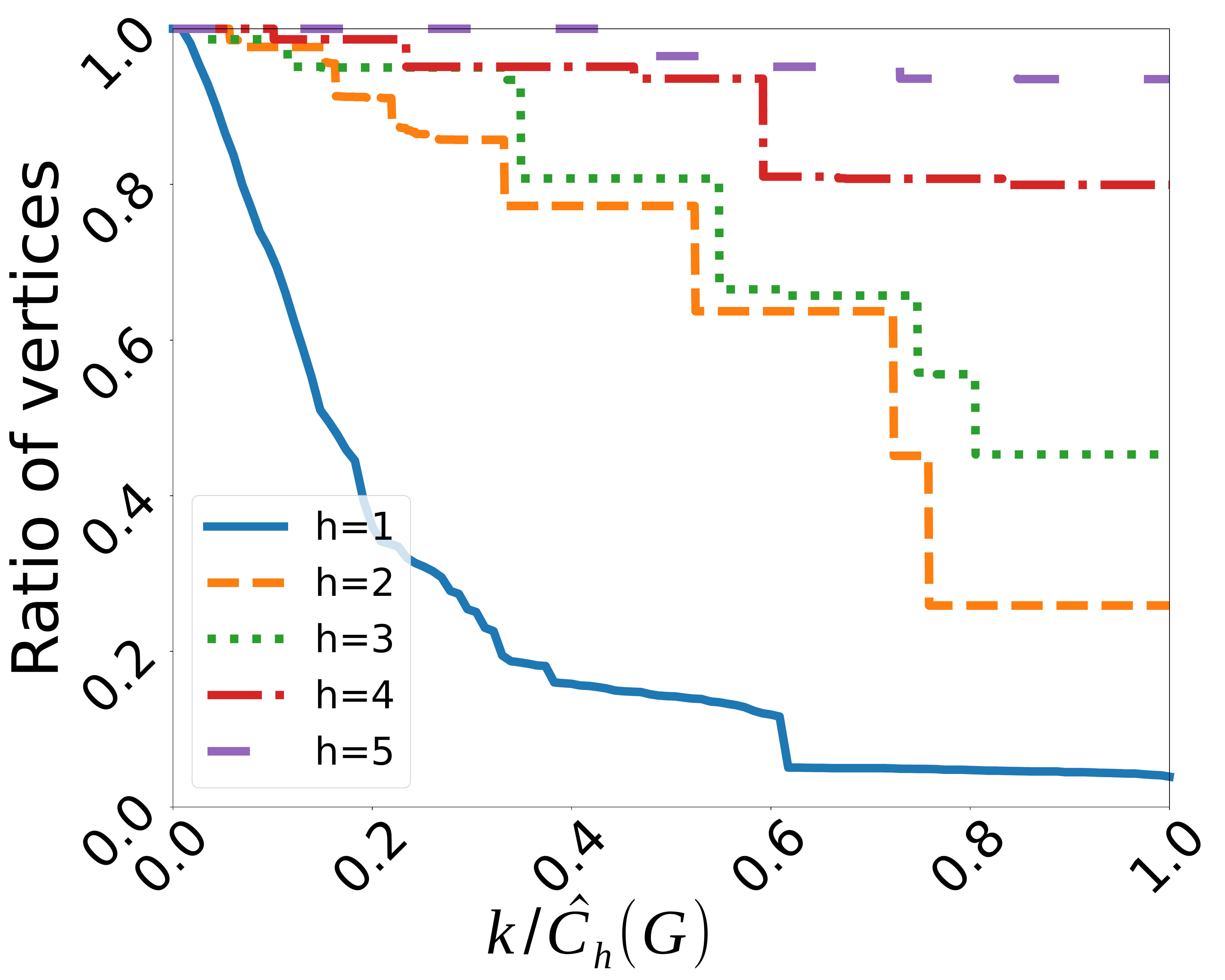}
\end{tabular}
\vspace{-3mm}
\caption
{For $1\leq h \leq 5$ how many vertices belong to the $(k,h)$-core $C_k$. On the $y$-axis we report $|C_k|/|V|$, while on the $x$-axis we report $k / \hat{C}_h(G)$. Here, $\hat{C}_h(G)$ denotes the largest value $k$ such that $G$ has a non-empty $(k,h)$-core.}
\label{Fig:core_distrib}

\vspace{2mm}

\begin{tabular}{cc}
\texttt{caAs} & \texttt{FBco} \\
\hspace{-4mm}\includegraphics[width=.245\textwidth]{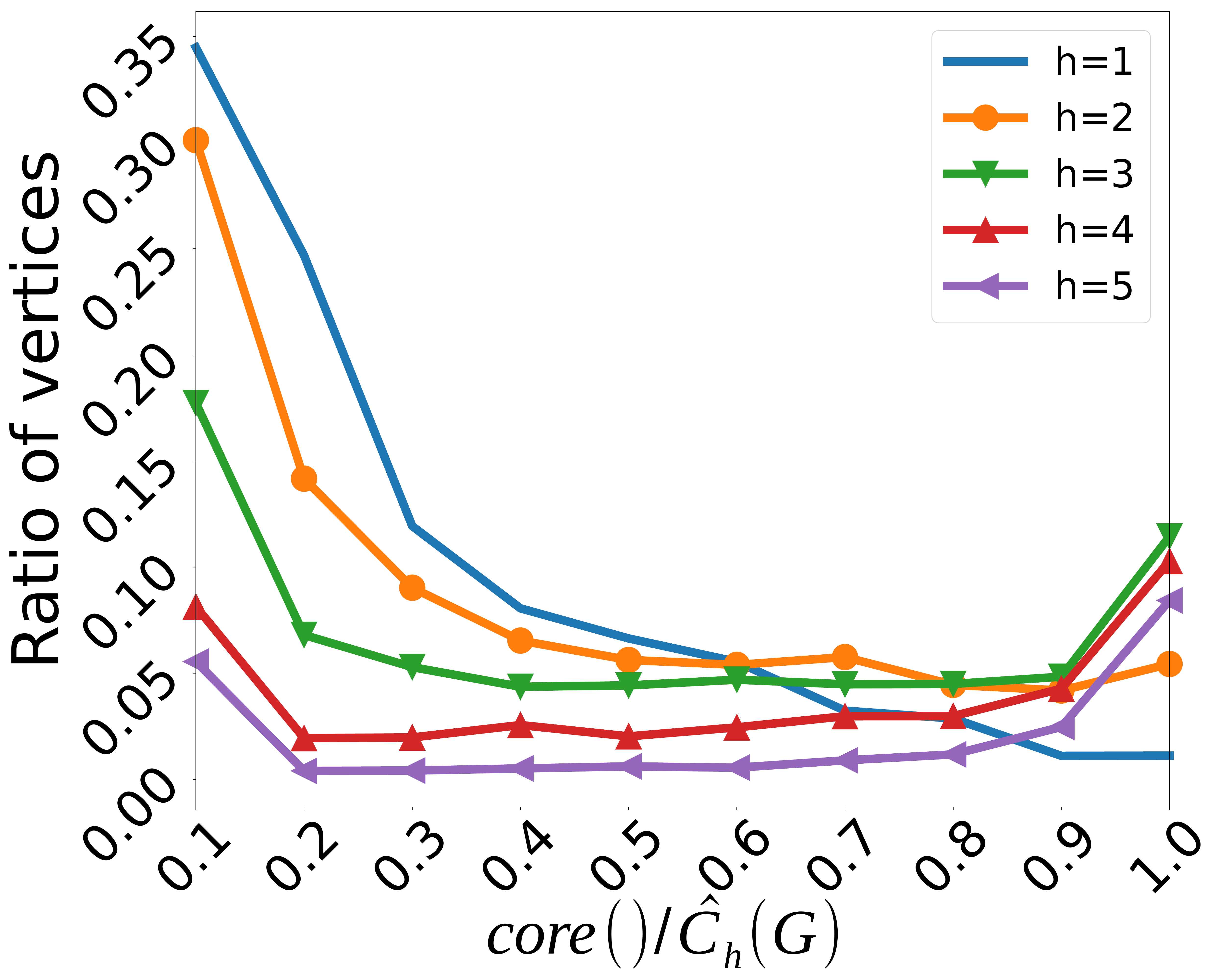}&
\hspace{-4mm}\includegraphics[width=.245\textwidth]{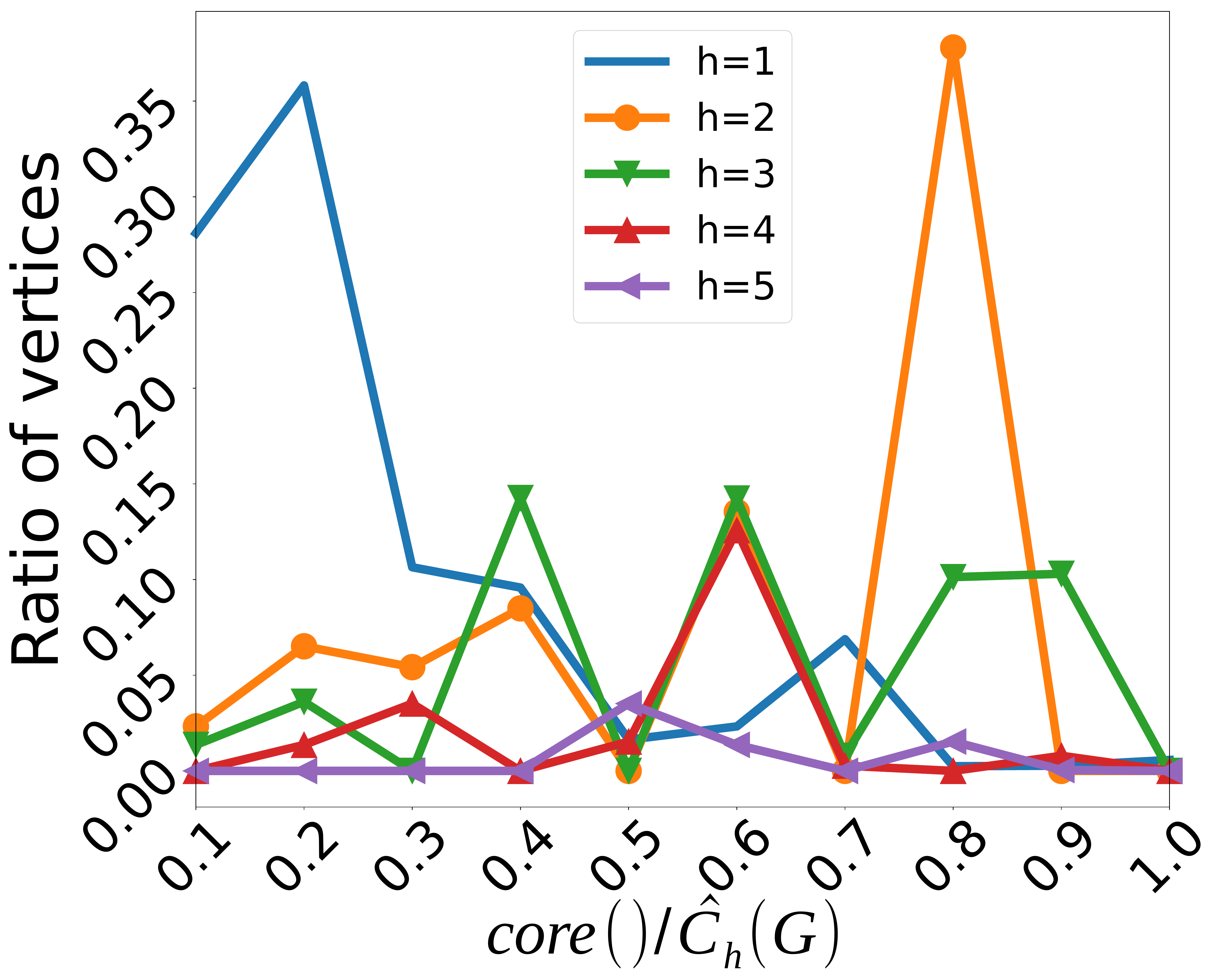}\\
\end{tabular}
\vspace{-3mm}
\caption{Fraction of vertices $v \in V$ having $core(v) = k$.
On the $x$-axis we report $k / \hat{C}_h(G)$ divided in ten intervals $(x_1,x_2],\ldots,(x_9,x_{10}]$: each point in the plot represents the fraction of vertices having $core() / \hat{C}_h(G)$ in $(x_i,x_{i+1}]$.}
\label{Fig:shell}
\vspace{-3mm}
\end{figure}

Our empirical characterization in Figure~\ref{Fig:core_distrib} and  \ref{Fig:shell} shows that the $(k,h)$-core index of a vertex for $h > 1$ provides very different information from the standard core index (i.e., $h = 1$). While there is no single value of $h$ which has more merits than the others, considering the core index for different values of $h$ (e.g., $h \in [1,4]$) might provide a more informative characterization of a vertex (a sort of ``spectrum'' of the vertex), than any core index alone.

Additional characterization experiments are presented in Appendix \ref{app:experiments}.

\subsection{Efficiency}
\label{subsec:exp2}
We next compare the runtime of the three algorithms described in Section~\ref{sec:algorithms}. In Table~\ref{tab:running_time} we report two measures: the runtime in seconds and the total number of point-to-point distance computations (or equivalently, the total sum of the sizes of all the $h$-BFS traversals executed).

\begin{table*}[t!]
	\centering	
	\caption{Running time (in seconds) and the number of computed point to point distances (i.e., the total number of possibly repeated vertices visited in all $h$-bfs). NT means that the algorithm did not terminate in 20 hours. In the two hardest networks (\texttt{sytb} and \texttt{hyves}) we report the time of $h$\textsf{-LB+UB} using 52 threads, while for all the other networks  we use single-threaded, sequential version of $h$\textsf{-LB+UB}.}
		\vspace{-2mm}
	\label{tab:running_time}
	    \setlength\tabcolsep{2pt}
\resizebox{\textwidth}{!}{
	\begin{tabular}{r|ccc|ccc|c|ccc|ccc|c|ccc|ccc|}

\multicolumn{1}{c}{} & \multicolumn{3}{c}{runtime (s)} & \multicolumn{3}{c}{visits $\times 10^{8}$} &
\multicolumn{1}{c}{} & \multicolumn{3}{c}{runtime (s)} & \multicolumn{3}{c}{visits $\times 10^{8}$} &
\multicolumn{1}{c}{} & \multicolumn{3}{c}{runtime (s)} & \multicolumn{3}{c}{visits $\times 10^{8}$} \\
		\cline{2-7} \cline{9-14} \cline{16-21}
		& $h=2$  & $h=3$ & $h=4$ & $h=2$  & $h=3$ & $h=4$ & $\;\;$ & $h=2$  & $h=3$ & $h=4$ & $h=2$  & $h=3$ & $h=4$ & $\;\;$ & $h=2$  & $h=3$ & $h=4$ & $h=2$  & $h=3$ & $h=4$ \\
		\cline{2-7} \cline{9-14} \cline{16-21}
\sf{$h$-BZ} & 3.72&269.34&380.85 & 0.87&28.91&33.68 & $\;$ $\;$&158.30&2825.41&14333.30	&14.55&232.88&1153.18& $\;$ $\;$ &283.63&16156.80&72332.70&55.95&	2032.47&6591.63 \\

$h$\textsf{-LB} & \textbf{0.17} & 1.19 & 1.50 &\textbf{0.06}&0.16&0.26 & $\;$ $\;$ &\textbf{0.95}& 128.16 &940.69& \textbf{0.13}&\textbf{10.67}&73.70 & $\;$ $\;$ &5.52&	560.20&	4835.06&1.06&75.19&	414.82 \\

$h$\textsf{-LB+UB} & 0.24 & \textbf{0.96} & \textbf{1.48} & 0.08&\textbf{0.13}&\textbf{0.25} & $\;$ $\;$ & 1.19&	\textbf{92.68} &\textbf{122.54}  &\textbf{0.13}&18.43&\textbf{8.65 }& $\;$ $\;$ & \textbf{5.17}&\textbf{91.39}&	\textbf{372.93} & 	\textbf{0.62}&\textbf{10.54}&\textbf{	32.81} \\
\cline{2-7} \cline{9-14} \cline{16-21}
\multicolumn{1}{c}{} & \multicolumn{6}{c}{\texttt{FBco}} &
\multicolumn{1}{c}{} & \multicolumn{6}{c}{\texttt{caHe}} &
\multicolumn{1}{c}{} & \multicolumn{6}{c}{\texttt{caAs}}\\
\multicolumn{21}{c}{}\\
		\cline{2-7} \cline{9-14} \cline{16-21}
				& $h=2$  & $h=3$ & $h=4$ & $h=2$  & $h=3$ & $h=4$ & $\;\;$ & $h=2$  & $h=3$ & $h=4$ & $h=2$  & $h=3$ & $h=4$ & $\;\;$ & $h=2$  & $h=3$ & $h=4$ & $h=2$  & $h=3$ & $h=4$ \\
		\cline{2-7} \cline{9-14} \cline{16-21}
\sf{$h$-BZ} &280.81&NT&NT &87.45&NT&NT& $\;$  &18.33&379.82&6451.33 &3.63& 81.36& 1275.23&  $\;$ &4.68&10.60&23.25& 0.36&1.24&3.48 \\

$h$\textsf{-LB} &\textbf{4.30}&1864.09&54762.10&  1.13&397.71&10989.5&  $\;$ & \textbf{2.51}&\textbf{29.27}&295.78  &\textbf{0.30}&4.70& 64.11& $\;$  &\textbf{3.18}&\textbf{6.75}&\textbf{11.47}& \textbf{0.25}&\textbf{0.66}&\textbf{1.64}\\

$h$\textsf{-LB+UB} & 6.76&\textbf{220.72}&\textbf{3556.72}&\textbf{1.06}&\textbf{33.96}&\textbf{636.52} & $\;$  & 12.98&	51.92&\textbf{190.88} & 0.59& \textbf{4.34}&\textbf{25.97} &  $\;$ &36.14&118.94&	139.80& 0.43&1.17&2.27 \\
\cline{2-7} \cline{9-14} \cline{16-21}
\multicolumn{1}{c}{} & \multicolumn{6}{c}{\texttt{doub}}&
\multicolumn{1}{c}{} & \multicolumn{6}{c}{\texttt{amzn}} &
\multicolumn{1}{c}{} & \multicolumn{6}{c}{\texttt{rnPA}} \\
\multicolumn{21}{c}{}\\
\cline{2-7} \cline{9-14} \cline{16-21}
		& $h=2$  & $h=3$ & $h=4$ & $h=2$  & $h=3$ & $h=4$ & $\;\;$ & $h=2$  & $h=3$ & $h=4$ & $h=2$  & $h=3$ & $h=4$ & $\;\;$ & $h=2$  & $h=3$ & $h=4$ & $h=2$  & $h=3$ & $h=4$ \\
\cline{2-7} \cline{9-14} \cline{16-21}
\sf{$h$-BZ} &5.74&13.26&27.10 &0.43&1.48&4.09& $\;$ & 154185.00 & NT & NT & 49035.00  & NT& NT & $\;$ & 56065.90  & NT& NT& 20493.07  & NT& NT \\
$h$\textsf{-LB} &\textbf{4.21}&\textbf{8.44}&\textbf{13.90}& \textbf{0.30}&\textbf{0.80}&\textbf{1.95}&  $\;$ &\textbf{102.75}&NT&NT& \textbf{33.36 } & NT&NT & $\;$  & \textbf{113.48 } & 42163.60&NT& \textbf{58.98 } &9467.16& NT \\
$h$\textsf{-LB+UB} &56.89&184.29&208.38 & 0.52&1.42&2.71 & $\;$ & 192.46  &\textbf{3192.07 }& \textbf{9310.85 }& 41.84  & \textbf{2085.06}& \textbf{7636.61 }& $\;$ & 440.93  &\textbf{3724.94}& \textbf{48038.70}& 76.69 &\textbf{2710.22}&\textbf{118834.25} \\
\cline{2-7} \cline{9-14} \cline{16-21}
\multicolumn{1}{c}{} & \multicolumn{6}{c}{\texttt{rnTX}} &
\multicolumn{1}{c}{} & \multicolumn{6}{c}{\texttt{sytb}} &
\multicolumn{1}{c}{} & \multicolumn{6}{c}{\texttt{hyves}} \\
\multicolumn{21}{c}{}\\
	\end{tabular}
}
\vspace{-4mm}
\end{table*}

We observe that, on every network, {\sf $h$\textsf{-LB}} and {\sf $h$\textsf{-LB+UB}} algorithms outperform the baseline {\sf$h$\textsf{-BZ}} algorithm in terms of running time. In all the cases, {\sf $h$\textsf{-LB}} and  {\sf $h$\textsf{-LB+UB}} reduce the number of computed distance pairs for at least one order of magnitude w.r.t. {\sf$h$\textsf{-BZ}} algorithm.
We also notice that {\sf $h$\textsf{-LB}} outperforms {\sf $h$\textsf{-LB+UB}} on road networks: this type of networks is in general sparse and exhibits a low $h$-degree even for the vertices belonging to the inner most cores.
When comparing  {\sf $h$\textsf{-LB}} with {\sf $h$\textsf{-LB+UB}} on other types of networks, we find that the former is often faster with $h=2$, while the latter is generally faster with $h>2$. This happens because as we increase $h$, the vertices in the inner most cores exhibit a larger $h$-degree and then, avoiding multiple re-computation of $h$-degree for such vertices, as the {\sf $h$\textsf{-LB+UB}} algorithm does, allows to speed up the decomposition.

\begin{table}[t!]
	\centering
	\caption{Relative error / fraction of vertices s.t the bound is tight (i.e. the bound is equal to the core index): on the left comparison between lower-bounds $LB_1$ and $LB_2$, on the right comparison between upper bound $UB$ and baseline $h$-degree.}
	\vspace{-2mm}
	\label{tab:error_bounds}
	\scriptsize
	\begin{tabular}{r|c|c|c|c|c|}		
		\multicolumn{1}{c}{}    & \multicolumn{1}{c}{$LB_1$}      & \multicolumn{1}{c}{$LB_2$} & \multicolumn{1}{c}{}    &\multicolumn{1}{c}{$h$-degree} & \multicolumn{1}{c}{$UB$}                     \\
		\cline{2-3} 	\cline{5-6}
	& 0.86 / 3.9\% & 0.35 / 19.2\% &  $h=2$ & 0.44 / 19.4\%& 0.01 / 53.6\% \\
		\texttt{caHe} & 0.95 / 3.8\%  & 0.78 / 4.4\%  &    $h=3$  & 0.40 / 10.3\%  & 0.01 / 29.8\% \\
		& 0.90 / 4.5\% & 0.42 / 6.1\%&   $h=4$   & 0.28 / 7.3\% & 0.01 / 17.9\%   \\	
		\cline{2-3} 	\cline{5-6}
		& 0.79 / 5.3\%  & 0.18 / 34.3\%  &  $h=2$ 	& 0.35 / 27.9\%  & 0.02 / 64.5\%   \\
		\texttt{caAs} & 0.92 / 5.1\%  &0.62 / 6.3\% & $h=3$& 0.32 / 15.1\%  & 0.01 / 57.2\%  \\
		& 0.87 / 6.5\%  & 0.31 / 9.5\%  & $h=4$& 0.37 / 11.3\%  & 0.01 / 26.4\%     \\
	\cline{2-3} 	\cline{5-6}
		& 0.69 / 2.1\%  & 0.09 / 56.5\%  &             $h=2$     & 0.45 / 16.1\%  & 0.01 / 81.4\%    \\
		\texttt{amzn} & 0.88 / 0.0\%  & 0.47 / 0.0\%  &  $h=3$   & 0.59 / 9.0\%  & 0.03 / 42.0\% \\
		& 0.81 / 0.1\%  & 0.33 / 12.7\% &   $h=4$ & 0.63 / 6.2\%  & 0.05 / 28.7\%    \\
		\cline{2-3} 	\cline{5-6}
		& 0.44 / 2.6\%  & 0.24 / 24.6\% &   $h=2$   & 0.59 / 20.3\%  & 0.01 / 98.2\%   \\
		\texttt{rnPA} 	& 0.71 / 0.1\%  & 0.58 / 0.1\% &    $h=3$   & 0.66 / 14.8\%  & 0.01 / 90.3\%   \\
		& 0.51 / 0.2\%  & 0.25 / 7.2\% &     $h=4$  & 0.70 / 9.0\%  & 0.01 / 79.9\%   \\
		\cline{2-3} 	\cline{5-6}
	\end{tabular}
\end{table}

\subsection{Lower and upper bounds}\label{subsec:expbounds}
We next asses on the effectiveness of the lower and upper bounds, showing their usefulness in reducing the computation, especially on the harder problem instances.

\spara{Lower bounds}.  Table~\ref{tab:error_bounds} reports the relative error with respect to the correct core index of each vertex and the fraction of vertices for which the bound is exactly the core index.

As expected the results confirm that, in general, $LB_2$ is tighter than $LB_1$ exhibiting a smaller relative error and an higher percentage of vertices for which the bound is tight (i.e. equal to the core index). Table \ref{tab:bounds_runtime} (left-hand side) reports the runtime comparison of the algorithms equipped with: no lower bound (Corresponding to algorithm $h$-\textsf{BZ}), $LB_1$
(corresponding to algorithm $h$-\textsf{LB} with $LB_1$ instead of $LB_2$), and $LB_2$ (corresponding to the standard $h$-\textsf{LB} algorithm). We can observe that the benefits on runtime of the lower bounds is usually one order of magnitude. Among the two lower bounds, the benefits of $LB_2$ over $LB_1$ increase with the complexity of the problem instance: for higher values of $h$, on the larger and denser networks, it becomes more visible (e.g., \texttt{amzn} in Table \ref{tab:bounds_runtime} for $h = 4$). On sparse networks, such as the road network \texttt{rnPA}, the computation is very fast and the overhead of computing $LB_2$ starting from $LB_1$ is no longer worth.

\begin{table}[t!]
	\centering
	\caption{Effect of bounds on running time (in seconds):
no lower bound (algorithm $h$-\textsf{BZ}), $LB_1$ (algorithm $h$-\textsf{LB} with $LB_1$ instead of $LB_2$),
 $LB_2$ (algorithm $h$-\textsf{LB}), $h$-degree (algorithm $h$\textsf{-LB+UB} with $h$-degree instead of $UB$), $UB$  (algorithm $h$\textsf{-LB+UB}).
}
	\vspace{-2mm}
	\label{tab:bounds_runtime}
	\scriptsize
	\begin{tabular}{r|c|c|c|c|c|c|}		
		\multicolumn{1}{c}{}   & \multicolumn{1}{c}{no $LB$}  & \multicolumn{1}{c}{$LB_1$}      & \multicolumn{1}{c}{$LB_2$} & \multicolumn{1}{c}{}    &\multicolumn{1}{c}{$h$-degree} & \multicolumn{1}{c}{$UB$}                     \\
		\cline{2-4} 	\cline{6-7}
                                  & 158.30 & 1.58 & \textbf{0.95} & $h=2$  & 1.87 & \textbf{1.19} \\
		\texttt{caHe}   & 2825.41 & 143.29 & \textbf{128.16} & $h=3$  & \textbf{23.45}& 92.68  \\
		                          & 14333.30 & 1229.54& \textbf{940.69} & $h=4$  & 308.91& \textbf{122.54}  \\	
		\cline{2-4} 	\cline{6-7}
                                    & 282.63 & 6.70 & \textbf{5.53} & $h=2$ & 6.39& \textbf{5.17} \\
		\texttt{caAs}   & 16156.80 & 590.45 & \textbf{560.20} & $h=3$ & 191.25& \textbf{91.39}  \\
		                          & 72332.70 & 5472.47 & \textbf{4835.06} & $h=4$ &1519.4& \textbf{372.93} \\	
		\cline{2-4} 	\cline{6-7}
                                & 18.33 & 3.30 & \textbf{2.51} & $h=2$ & 32.99 & \textbf{12.98} \\
		\texttt{amzn}   & 379.82 & 34.91 & \textbf{29.27} & $h=3$  & 89.71& \textbf{51.92}  \\
		                          & 6451.33 & 529.84 & \textbf{295.78} & $h=4$ & 404.80 & \textbf{190.88} \\	
		\cline{2-4} 	\cline{6-7}
                        & 4.68 & \textbf{3.00} & 3.18& $h=2$  & 36.64 & \textbf{36.14} \\
		\texttt{rnPA}   & 10.60 & \textbf{5.98} & 6.75&$h=3$ & 124.26 & \textbf{118.94} \\
		                & 23.25 & 11.97 &\textbf{11.47}& $h=4$ & 143.71& \textbf{139.80}  \\	
		\cline{2-4} 	\cline{6-7}
	\end{tabular}
\end{table}

For what concerns the definition of $LB_2$, we also confirmed that considering the $h/2$-neighborhood of a vertex is more effective than considering the 1-neighborhood. For instance, in \texttt{caHe} for $h=4$, using algorithm $h$\textsf{-LB+UB}: despite the fact that
computing  $LB_2$  considering the $h/2$-neighborhood requires 0.3 seconds more than  considering the $1$-neighborhood, using the latter leads to an increase of the running time of 961 seconds since it computes $7.6\times 10^8$ point-to-point distances more.

Finally, assessing the benefits of $LB_3$ in isolation is complicated by the fact that this bound is deeply entangled in the logic of Algorithm $h$\textsf{-LB+UB}: it
is dynamically recomputed each time the algorithm moves to a new partition, getting closer and closer to the actual value of the core index of each vertex, as the computation progresses to higher partitions. Therefore, it depends greatly on the way the partitions are defined and, indirectly, it depends on the upper bound.

\spara{Upper bound}. For what concerns the upper bound $UB$ used in algorithm $h$\textsf{-LB+UB} we compare it with a simple upper-bound for the core index of a vertex, namely its $h$-degree. Table~\ref{tab:error_bounds} (right-hand side) shows that the upper bound $UB$ is way more accurate than the baseline, often very close to the actual value of the core index.  Table \ref{tab:bounds_runtime}  (right-hand side) reports the runtime comparison for algorithm $h$\textsf{-LB+UB} equipped with $UB$  or when the upper bound is substituted with $h$-degree: we can observe  that, as discussed for the lower bounds, the benefits of $UB$ becomes more evident on harder problem instances (e.g., \texttt{amzn} in Table \ref{tab:bounds_runtime} for $h = 4$).

In conclusion, all introduced bounds are effective towards achieving scalability: the cost of computing them is highly recompensed in saved computation, especially on harder problem instances (i.e., high value of $h$ on larger denser networks).

\begin{figure}[t!]
\vspace{-4mm}
	\caption{Runtime of $h$\textsf{-LB+UB} algorithm (52 threads) on subgraphs of different size sampled from \texttt{lj} network.}
\vspace{-2mm}
	\centering
\includegraphics[width=.3\textwidth]{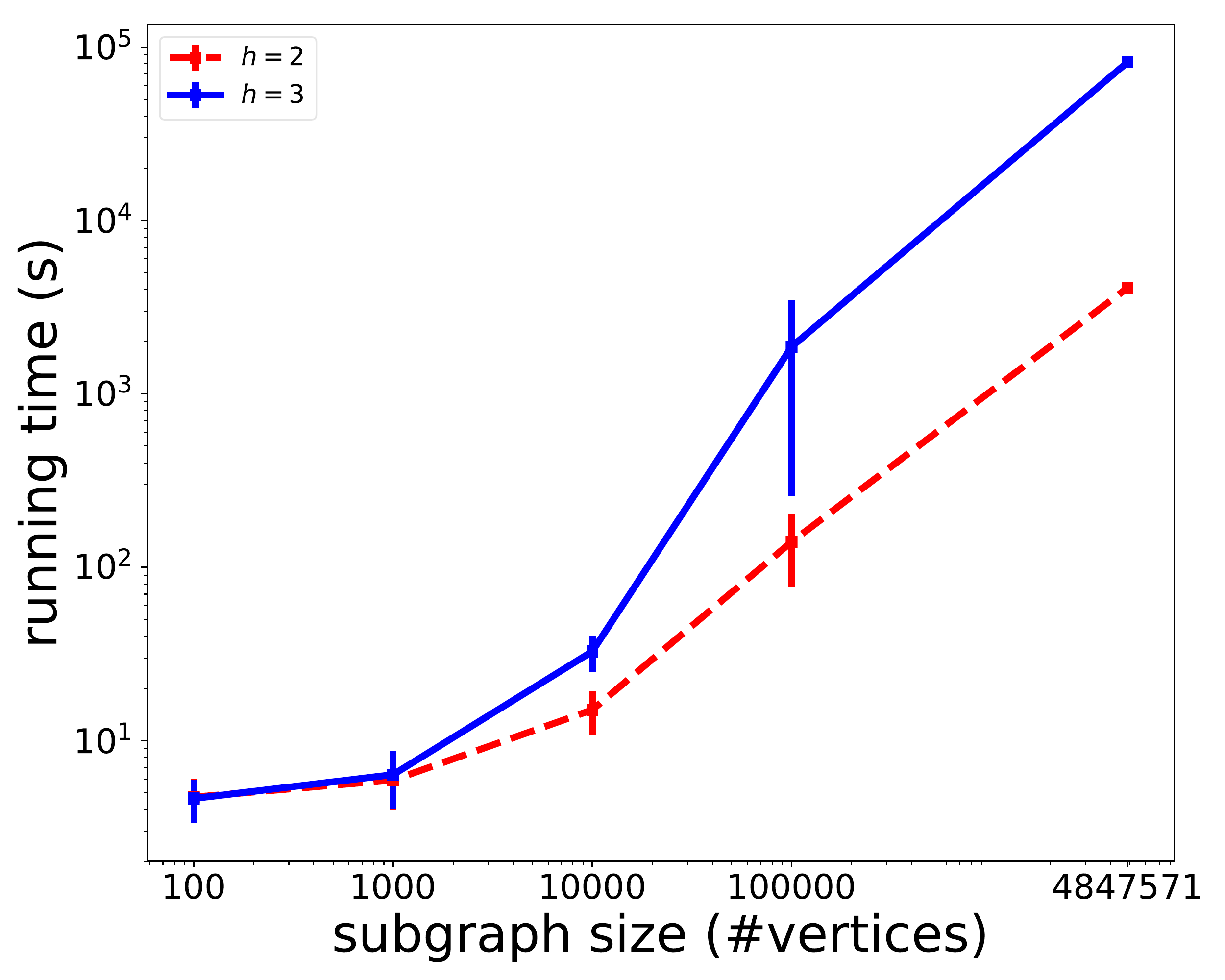}\\
	\label{fig:scalability}
\vspace{-2mm}
\end{figure}

\subsection{Scalability}\label{subsec:expscalability}
We next aim at $(i)$ showing that our best performing algorithm,  $h$\textsf{-LB+UB}, can scale to compute the distance-generalized core decomposition on a large and dense network, and $(ii)$ study how the runtime grows with the growth of the network.
For our purposes we use the \texttt{lj} network and we sample intermediate subgraphs of different sizes.
Each subgraph is generated by means of \emph{snowball sampling}: we select a random vertex from the original network and we run a BFS from it stopping as soon as we have visited $|V'|$ vertices and we return the subgraph $G'$ induced by those vertices. We select $|V'|$ equal to 100,1000,10000 and 100000. For each size, given the stochastic selection of the seed vertex, we sample 10 different subgraphs (except, of course for the experiment using the whole \texttt{lj} network).

Figure~\ref{fig:scalability} reports the average running time (and the standard deviation) for each sample size, using 52 threads. We notice that with $h=2$ the algorithm exhibits an almost linear scalability and we can compute the complete decomposition of \texttt{lj} network in one hour.
For $h = 3$ the runtime is similar to $h = 2$ for the subgraphs with $|V'|\leq 10000$, while for larger graphs the computation becomes more demanding.

\subsection{Application: maximum $h$-club problem}
\label{subsec:exp3}
We next compare Algorithm \ref{alg:h-club} that we proposed in
\S\ref{subsec:club} for maximum $h$-club problem, with the state-of-the-art {\sf DBC} and {\sf ITDBC}~\cite{MB15}  algorithms based on linear programming: the software obtained from the authors of~\cite{MB15},  is implemented in \texttt{C++} and uses Gurobi\footnote{\url{http://www.gurobi.com/}} optimizer 7.5.1 to solve the IP formulations. We set up Gurobi to use a single core.

Table~\ref{tab:kclubs} reports the running time of Algorithm \ref{alg:h-club}
including the time needed to compute the $(k,h)$-core decomposition.

Solving the linear program on much smaller graphs, Algorithm~\ref{alg:h-club} (using either DBC or ITDBC as black-box algorithm) is, in our experiments and set up, much faster
than {\sf DBC}  (which solves liner program on the entire input graph) and the iterative approach of {\sf ITDBC}.
For the same reason Algorithm~\ref{alg:h-club} requires much less memory than {\sf DBC}, being able to solve the maximum $h$-club
problem also in larger graphs where {\sf DBC} fails due to the excessive size of linear program.

\begin{table}[h!]
	\vspace{-2mm}
	\centering
	\caption{Runtime (seconds) for the maximum $h$-club problem (NT = not terminate within 24 hours; OM = requires more than 128GB RAM).}
	\vspace{-4mm}
	\label{tab:kclubs}
	\small
	\setlength\tabcolsep{2pt}
	\begin{tabular}{r|c|cccc|l}		
\multicolumn{1}{c}{}    & \multicolumn{1}{c}{Size of }      & \multicolumn{1}{c}{} &\multicolumn{1}{c}{} & \multicolumn{1}{c}{Alg. \ref{alg:h-club} +} &         \multicolumn{1}{c}{Alg. \ref{alg:h-club} +} &               \\
	\multicolumn{1}{c}{}    & \multicolumn{1}{c}{max $h$-club}      & \multicolumn{1}{c}{DBC} &\multicolumn{1}{c}{ITDBC} & \multicolumn{1}{c}{DBC} &     \multicolumn{1}{c}{{ITDBC}} &                 \\ \cline{2-6}
		 &   1046   &23.9& 0.6&{\bf 0.18}     & \textbf{0.2}&   $h=2$ \\
		\texttt{FBco}&  1830    &187.7&55.1 & {\bf 12.1}    &  12.4   &       $h=3$       \\
		&3229&   51.7   &52.7&   {\bf 36.9}   & 37.1  &     $h=4$      \\ \cline{2-6}
		  &  512    &2517.1&485 &   {\bf 165.7}  & 588.8   &      $h=2$  \\
		\texttt{caHe} & 2268     &6056.9&20898 &  {\bf 355.9}    & \textbf{355.9}  &    $h=3$  \\
		&   NT   &NT&NT & NT      & NT &     $h=4$  \\ \cline{2-6}

		&   550   &OM&642 &     {\bf 2.5}  & \textbf{ 2.5}   &              $h=2$       \\
		\texttt{amzn} &  621    &OM&677&  {\bf 29.3}   & \textbf{29.3}   &   $h=3$  \\
		&  1397    &OM&636& {\bf 190.9}    &\textbf{ 190.9 }  &     $h=4$  \\ \cline{2-6}
				  	
				  	&   10   &OM&16382&   {\bf 4.2}  &    {\bf 4.2}   &           $h=2$       \\
				\texttt{rnTX} & 15     &OM&14420 &{\bf 8.4} & {\bf 8.4}  &     $h=3$          \\
				& 29     &OM&14601 & {\bf 13.9}  &  {\bf 13.9}    &   $h=4$   \\ \cline{2-6}
					 & 13     &OM&12238 &   {\bf 3.2}  & {\bf 3.2}    &        $h=2$   \\
						\texttt{rnPA} 	& 21     &OM&59539 & 128.3    &  {\bf 6.8}  &    $h=3$   \\
					&   29   &OM& 8195.8&  {\bf 11.5} &  {\bf 11.5}   &     $h=4$  \\ \cline{2-6}
					
	\end{tabular}
\vspace{-3mm}
\end{table}

\subsection{Application: landmarks selection for shortest path distance estimation }\label{subsec:landmarks}
Landmarks-based indexes play an important role in point-to-point (approximate) shortest-path query~\cite{sommer2014shortest,Potamias:2009,akiba2013fast}.
The idea is as follows.
Given a graph $G = (V,E)$ a pair of vertices $s,t \in V$ and another vertex $u \in V$  which plays the role of \emph{landmark}, we can approximate the shortest path distance $d_G(s,t)$ from the distances $d_G(s,u)$ and $d_G(u,t)$ by means of the following inequalities:
$$
d_G(s,t) \leq d_G(s,u) + d_G(u,t)
$$
$$
d_G(s,t) \geq |d_G(s,u) - d_G(u,t)|
$$
Therefore, if we are given a set $L \subseteq V$ of $\ell = |L|$ landmarks we can bound the distance $d_G(s,t)$ as follows:
$$
\max_{u \in L} |d_G(s,u) - d_G(u,t)| \leq d_G(s,t) \leq \min_{u \in L} d_G(s,u) + d_G(u,t)
$$
In the following we use $LB(s,t) = \max_{u \in L} |d_G(s,u) - d_G(u,t)|$ to denote the lower-bound and
$UB(s,t) = \min_{u \in L} d_G(s,u) + d_G(u,t)$ to denote the upper-bound.

The quality of the approximation of $d_G(s,t)$ obtained by means of these bounds is high when the vertices $s$ and $t$ have short distance to some landmarks. Therefore, when selecting landmarks we aim at ``covering'' as many vertices in the network as possible, by some landmark within a short distance. Our hypothesis is that vertices in the high cores of a distance-generalized core decomposition are good candidates to become landmarks, because they are part of a dense and large subgraph, containing many vertices all within a close distance, so that they are likely to be rather close to a very large portion of the network.

We next test our hypothesis. In particular we select  $\ell = 20$ landmarks at random from the core of maximum index (i.e., the $(k,h)$-core such that there is no nonempty core $(k',h)$-core with $k' > k$), and we do this for different values of $h \in [1,4]$. We compare against top-$\ell$ \emph{closeness centrality} ($cc$) vertexes, which is one of the best-performing heuristics in practice~\cite{Potamias:2009}, top-$\ell$ \emph{betweenness centrality} ($bc$) vertices and top-$\ell$ high $h$-degree (i.e., $deg^h_G$ for $1\leq h \leq 4 $) vertices.

After having selected  $\ell = 20$ landmarks we measure the approximation error for 500 randomly sampled couples of vertices $s,t \in V$. In particular, we report the relative error that we achieve by approximating $d_G(s,t)$ with the median of the two bounds:
$$
\left| \frac{LB(s,t) + UB(s,t)}{2} - d_G(s,t) \right| / d_G(s,t).
$$

As there is stochastic component, we perform each experiment 10 times and report the average results.

Table \ref{tab:landmarks} reports the results over medium-sized datasets for approximation error (the smaller the better).
The results are consistent across experiments, and consistent with the values  of approximation error in the literature. We can observe that selecting landmarks according to the distance-generalized core decomposition with $h >1$, produces very good landmarks, especially for $h = 4$, which are clearly outperforming the case $h =1$ and the other baselines. We can also observe that, while selecting the landmarks from the maximum $(k,h)$-core the accuracy increases for larger values of $h$, the same does not happen when selecting by higher $h$-degree.

\begin{table}[t!]
	\centering
	\caption{Landmarks selection. Approximation error selecting  randomly 20 vertices from the maximum $(k,h)$-core ($h \in [1,4]$), the top-20 by closeness ($cc$) or betweenness ($bc$) centralities, the  top-20 by  $deg^h_G$. For completeness sake in the bottom table we report maximum core index / number of vertices in that core. \label{tab:landmarks}}
	\vspace{-3mm}
\begin{small}
	\begin{tabular}{r|cccc|}
\multicolumn{1}{c}{} &	\texttt{FBco} & 	\texttt{caHe} &	\texttt{caAs} &	\multicolumn{1}{c}{\texttt{doub}} \\
\cline{2-5}
$h = 1$ &	 0.25	& 0.22 &	 0.18 &	 0.2 \\
$h = 2$ &	 0.16 	& 0.18 	&0.16 &	0.2 \\
$h = 3$ &	 0.12 	& 0.17 	& \textbf{0.14} 	& 0.17 \\
$h = 4$ &	\textbf{0.07} &	 \textbf{0.14} 	& \textbf{0.14} &	 \textbf{0.14} \\ \cline{2-5}
$cc$ &	0.26 &	 0.24  &	 0.22 	& 0.2 \\
{$bc$} &	{0.29} &	{0.21}&	{0.21}&	{0.26} \\ \cline{2-5}
{$deg^1_G$ }&{0.22}&	{0.23}&	{0.22}&	{0.26} \\
{$deg^2_G$} &{0.27}&	{0.23}&	{0.22}	&{0.26} \\
{$deg^3_G$} &	{0.28}&	{0.23}&	{0.22}	&{0.26} \\
{$deg^4_G$} &{0.26} &{0.23}&	{0.22}&{0.26} \\
		\cline{2-5}
	\end{tabular}
\end{small}

\smallskip  \smallskip

\begin{scriptsize}
\begin{tabular}{r|cccc|}
\multicolumn{1}{c}{} &	\texttt{ FBco} & 	\texttt{ caHe} &	\texttt{ caAs} &	\multicolumn{1}{c}{\texttt{ doub}} \\
\cline{2-5}
$h = 1$ &  115/158  &  238/239 &  56/57  &  15/1857\\
$h = 2$ &
 1045/1046  &  654/883 &  680/1741  &  423/2404\\
$h = 3$&
 1829/1830  &  2267/2268 &  4305/5898  &  4077/7071\\
$h = 4$ &   3228/3229  &  4392/5331 &  10252/11333  &  21460/41968\\
\cline{2-5}
	\end{tabular}
\end{scriptsize}
	\vspace{-4mm}
\end{table}

\section{Conclusions and Future Work}
\label{sec:conclusions}
In this paper we introduce the distance-generalized core decomposition and show that it generalizes many of the nice properties of the classic core decomposition, e.g., its connection with the notion of \emph{distance-generalized chromatic number}, or  its usefulness in speeding-up or approximating distance-generalized notions of dense structures, such as \emph{$h$-club} or the (distance-generalized) \emph{densest subgraph} and \emph{cocktail party} problems.
Some of these applications of the distance-generalized core decomposition stand as contributions \emph{per se}. In particular, our simple idea of using the $(k,h)$-core decomposition as a wrapper around any existing method for maximum $h$-club, is  shown empirically to provide important execution-time benefits, progressing beyond the state of the art of this active research topic.

This paper opens several future directions.
Our empirical characterization shows that the $(k,h)$-core index of a vertex for $h > 1$ provides very different information from the standard core index (i.e., $h = 1$).
Considering the core index for different values of $h$ (e.g., $h \in [1,4]$) might provide a more informative characterization of a vertex (a sort of ``spectrum'' of the vertex), than any core index alone. Investigating the properties of this ``spectrum'' of a vertex is worth further effort. Related to this, it is interesting to develop algorithms that for a given input graph would compute the $(k,h)$-core decompositions for different values of $h$ all at once.


\pagebreak

\appendix

\section{Proofs}
\label{sec:appendix}
\paragraph{Proof of Property~\ref{prop:unique}}
\begin{proof}
	By contradiction. Assume there exist two different
	$(k,h)$-cores $G[S]=(S,E[S])$ and $G[T]=(S,E[T])$ of $G$.
	Consider the subgraph $G[S\cup T]$ induced by the union of $S$ and $T$. It is straightforward that in such subgraph, every
	vertex has at least $k$ neighbors within distance $h$. Thus, also $G[S\cup T]$ is a $(k,h)$-core. It follows that both $G[S]$ and $G[T]$ are not maximal, and thus are not $(k,h)$-cores, contradicting the initial assumption.
\end{proof}
\paragraph{Proof of Property~\ref{prop:subgraph}}
\begin{proof}
	The property follows from the definition of  $(k,h)$-core through the following two observations:
	(1) in a subgraph in which every vertex has at least $k+1$
	neighbors within distance $h$,  every
	vertex also has at least $k$ neighbors within distance $h$ in that subgraph;
	(2) as we enlarge the subgraph and add more induced edges, the distance between two vertices may only decrease.
\end{proof}
\paragraph{Proof of Observation~\ref{th:lb1}}
\begin{proof}
	First, we prove that each ${\lfloor\frac{h}{2}\rfloor}$-neighbor $u_i$ of $v$ have $core(u_i)\geq {\lfloor\frac{h}{2}}\rfloor$.
	Recall that $h>1$. For all $u_i$,  $d_{G[V]}(v,u_i)\leq {\lfloor\frac{h}{2}\rfloor}$. It easy to see that, by triangle inequality,
	for all $u_i,u_j$ in the the ${\lfloor\frac{h}{2}\rfloor}$-neighborhood of $v$, we have $d_{G[V]}(u_j,u_i)\leq \lfloor\frac{h}{2}\rfloor + \lfloor\frac{h}{2}\rfloor \leq h$.
	This means that every $u_i$ has at least $deg^{{\lfloor\frac{h}{2}}\rfloor}_{G[V]}(v) - 1$ neighbors at distance $\leq h$ and one at distance $\leq \lfloor\frac{h}{2}\rfloor$
	(the vertex $v$ itself). By definition of $(k,h)$-core, each $u_i$ has at least $core(u_i)\geq deg^{{\lfloor\frac{h}{2}}\rfloor}_{G[V]}(v)$. Since $v$ has
	$deg^{{\lfloor\frac{h}{2}}\rfloor}_{G[V]}(v)$ neighbors with core $\geq deg^{{\lfloor\frac{h}{2}}\rfloor}_{G[V]}(v)$, then by definition of $(k,h)$-core,
	it is in the $deg^{{\lfloor\frac{h}{2}}\rfloor}_{G[V]}(v)$-core.
\end{proof}
\paragraph{Proof of Observation~\ref{th:lb2}}
	\begin{proof}
		Suppose that there exists at least a vertex $u$ such that $d_{G[V]}(u,v)\leq {\lceil\frac{h}{2}\rceil}$ and $LB_1(u) > LB_1(v)$. By definition of $LB_1$, this means that $u$ has at least $LB_1(u)$ neighbors $w_j$ at distance at most  $\lfloor\frac{h}{2}\rfloor$. However, since $d_{G[V]}(u,v)\leq \lceil\frac{h}{2}\rceil$, it is easy to see that $d_{G[V]}(w_j,v)\leq {h}$, for each $w_j$. This means that $v$ has at least $LB_1(u)$ neighbors at distance $\leq {h}$ and one vertex at distance $\leq \lceil\frac{h}{2}\rceil$ (the vertex $u$ itself), with core index bounded by $LB_1(u)$ (every $w_j$ is at distance at most $h$ from each other, since they are at distance $\lfloor\frac{h}{2}\rfloor$ from $u$; thus every $w_j$ has core index at least $LB_1(u)$), and so $LB_2(v)=LB_1(u)\leq core(v)$.
	\end{proof}

\paragraph{Proof of Observation~\ref{obs}}
\begin{proof}
	Suppose that there exists a vertex $v$ in a $(k,h)$-cores with $k \geq i$ s.t. $v\notin V[i]$, that means $UB(v)<i$. However, since $UB(v)$ is an upper bound on $core(v)$, we have $UB(v)\geq core(v)\geq k \geq i$, that is a contradiction.
\end{proof}

\paragraph{Proof of Property~\ref{prop:mindegree}}
\begin{proof}
	Let $d^*=\min(deg^h_{G[V']}(v)| v\in V')$.
	Clearly, every vertex $u \in V'$ must have $h$-degree at least $d^*$ in the subgraph $G[V']$.
	By the definition of $(k,h$)-core, every $u \in V'$ must be in $(d^*, h)$-core of $G[V']$.
	However, as $G[V']$ is a subgraph of $G$, it holds, for every vertex $u \in V'$, that:
	
	\noindent $core(u)\geq core_{G[V']}(u) \geq  d^*=\min(deg^h_{G[V']}(v)| v\in V').$
\end{proof}
\paragraph{Proof of Theorem~\ref{th:chromatic}}
\begin{proof}
	 We need to produce a distance-$h$ coloring such that the maximum number of colors required for that greedy approach is $1+\hat{C}_h(G)$. Since
	the distance-$h$ chromatic number is upper bounded by the
	maximum number of colors required for  a distance-$h$ coloring,
	then the theorem follows.
	
	We perform a greedy distance-$h$ coloring by starting with an empty graph and adding vertices in the reversed order w.r.t. the one of the ``peeling'' algorithm for core decomposition (e.g., Algorithm~\ref{alg:peeling_algo}), which iteratively removes the vertex having the smallest $h$-degree in the current
	subgraph.
	
	If we consider such a reverse order, when a vertex is added, it can have at most $k$ $h$-neighbors,
	where $k$ is its core index. This holds because the core index of a vertex is the maximum value between the $h$-degeneracy of the current subgraph and the vertex's $h$-degree in the current subgraph (this property is used also by Algorithm~\ref{alg:peeling_algo}, line 10).
	Therefore, to color the subgraph till the addition of
	that vertex, we need at most $(k+1)$ different colors. Since the maximum core-index of a vertex is the $h$-degeneracy $\hat{C}_h(G)$ of
	the input graph $G$, one may need up to $\hat{C}_h(G)+1$ colors with this greedy $h$-coloring approach. Hence,
	the distance-$h$ chromatic number, $\chi_h(G) \le 1+\hat{C}_h(G)$.
\end{proof}
\paragraph{Proof of Theorem~\ref{th:core_club}}
\begin{proof}
	We shall prove by contradiction. Let, if possible, $C$ be an $h$-club of size $k+1$, such that $C=C_1\cup C_2$,
	$C_1\cap C_2 = \emptyset$, and $C \cap \mathcal{C}(k,h)=C_1$. Clearly, $C\setminus \mathcal{C}(k,h)=C_2$.
	Let us consider that $\mathcal{C}' =\mathcal{C}(k,h)\cup C_2$.
	It is easy to verify that $\mathcal{C}'$ is a $(k,h)$-core of $G$. This is because every vertex in $C_2$ must
	have a path of length at most $h$, to $k$ other vertices of $\mathcal{C}'$, and all these paths will pass through
	vertices inside $C \subseteq \mathcal{C}'$. It follows from the definition of $h$-club.
	However, this is a contradiction, unless $C_2=\emptyset$, since our assumption
	was that $\mathcal{C}(k,h)$ is the $(k,h)$-core of $G$, and $\mathcal{C}(k,h)\subseteq {\mathcal C}'$. 
\end{proof}
\paragraph{Proof of Theorem~\ref{th:densest}}
\begin{proof}
	We assume that $G[S^*]$ is the distance-$h$ densest subgraph.
	Let $v \in S^*$, $|S^*|=s^*$. Then, for all $v_1 \in S^*$
	\begin{align}
	f_h(S^*) = \frac{\sum_{v\in S^*} deg^h_{G[S^*]}(v)}{s^*} \ge  \frac{\sum_{v\in S^*\setminus\{v_1\}} deg^h_{G[S^*\setminus\{v_1\}]}(v)}{s^*-1}
	\label{ineq:dense1}
	\end{align}
	The above holds because $G[S^*]$ is the distance-$h$ densest subgraph. Also, assuming a connected graph, $s^*>1$.
	Now, it is easy to verify the following.
	\begin{small}
		\begin{align}
		&\sum_{v\in S^*\setminus\{v_1\}} deg^h_{G[S^*\setminus\{v_1\}]}(v) \ge \sum_{v\in S^*} deg^h_{G[S^*]}(v) - 2\left[deg^h_{G[S^*]}(v_1)\right] \nonumber & \\
		& \qquad \qquad - deg^h_{G[S^*]}(v_1)\left[deg^h_{G[S^*]}(v_1)-1\right] \nonumber & \\
		& = \sum_{v\in S^*} deg^h_{G[S^*]}(v) - deg^h_{G[S^*]}(v_1) - \left[deg^h_{G[S^*]}(v_1)\right]^2 &
		\label{ineq:dense2}
		\end{align}
	\end{small}
	In the second line of the above inequality, we subtract $2[deg^h_{G[S^*]}(v_1)]$, because we consider
	the set $S^*\setminus\{v_1\}$, i.e., vertex $v_1$ is removed from $S^*$. Furthermore, we subtract $deg^h_{G[S^*]}(v_1)[deg^h_{G[S^*]}(v_1)-1]$,
	because by deleting $v_1$, every vertex in $N_{G[S^*]}(v_1,h)$ can be disconnected (or, their distance
	may increase from ``$\le h$'' to ``$>h$'') from at most every other
	vertex in $N_{G[S^*]}(v_1,h)$.
	Next, by combining Inequalities~\ref{ineq:dense1} and \ref{ineq:dense2}, one can derive, for all $v_1\in S^*$:
	\begin{align}
	deg^h_{G[S^*]}(v_1) \ge (\sqrt{f_h(S^*)+0.25}-0.5)
	\label{eq:density}
	\end{align}
	Consider now any of our $(k,h)$-core decomposition algorithms (e.g., for simplicity,
	let us consider the $h$\textsf{-BZ} method in Algorithm~\ref{alg:peeling_algo},
	which continues by peeling the vertex having the smallest $h$-degree in the current
	subgraph). Let $W \subset V$ be the last $(k,h)$-core found before
	the first vertex  $u$ that belongs to $S^*$ is removed.
	Clearly, $S^* \subseteq W$. Due to the greediness of the algorithm,
	the following holds: $$\forall w \in W \; deg^h_{G[W]}(w) \ge deg^h_{G[W]}(u).$$
	Moreover, since $S^* \subseteq W$ it holds that
	$$deg^h_{G[W]}(u) \ge deg^h_{G[S^*]}(u).$$
	\vspace{-2mm}
	Hence,
	\vspace{-2mm}
	$$\sum_{w\in W} deg^h_{G[W]}(w) \ge |W|\left[deg^h_{G[S^*]}(u)\right];$$
	$$\frac{\sum_{w\in W} deg^h_{G[W]}(w)}{|W|} \ge deg^h_{G[S^*]}(u).$$
	Substituting in Equation~\ref{eq:density}, we get:
	$$
	f_h(W) \ge (\sqrt{f_h(S^*)+0.25}-0.5)
	$$
\end{proof}

\section{Distance-generalized cocktail party}\label{subsec:cocktail}
The \emph{community search problem} has drawn a lot of attention in the last years \cite{SozioLocalSIGMOD14,KtrussSIGMOD14,ruchansky2015minimum,fang2017attributed,huang2017attribute,HuangLX17}: given a set of query vertices the problem requires to find  a subgraph that contains the query vertices and that is densely connected.
One of the first formulations of this problem, known as \emph{cocktail party} (from the title of the paper), was by Sozio and Gionis~\cite{sozio2010}, which adopted the  minimum degree within the subgraph as the density measure to be maximized.

We adopt the definition by Sozio and Gionis~\cite{sozio2010} and study its distance-generalization
by considering the minimum $h$-degree as a measure of how well-connected is the subgraph.
\begin{problem}[Distance-generalized cocktail party]
	Given a graph $G=(V,E)$, a set of query vertices $Q\subseteq V$,  and a distance threshold $h \in \mathbb{N}^+$,  find a set of vertices $S^*  \subseteq  V$ such that it contains $Q$,  it is connected and it  maximizes the minimum $h$-degree:
$$
 S^*=\underset{Q \subseteq S\subseteq V}{\argmax}\min_{v\in S} deg^h_{G[S]}(v)
$$
\end{problem}
It is straightforward to see that an optimal solution to this problem is given by the $(k,h)$-core with the largest $k$
containing all the vertices $Q$ and in which all the vertices $Q$ are connected.
To solve this problem efficiently we can
adapt $h$\textsf{-LB+UB} (Algorithm~\ref{alg:topdown}), which finds higher cores earlier. More in details, consider the \texttt{for} loop in Lines~\ref{alg:topdown:parpart:start}--\ref{alg:topdown:parpart:end}, Algorithm~\ref{alg:topdown},  and consider the first iteration such that all the query vertices $Q$ have their core index assigned to $k_{min}$.
If all the query vertices are connected in the $(k_{min},h)$-core we return the respective connected component.
Otherwise, we iteratively decrease $k_{min}$ and assign the core index to the corresponding vertices until all the query vertices belong to the same connected component in a $(k_{min},h)$-core. 

\section{ADDITIONAL EXPERIMENTS}
\label{app:experiments}

\begin{figure}[t!h!]
\vspace{-3mm}
\begin{minipage}{0.23\textwidth}
\centering
\includegraphics[width=.95\linewidth]{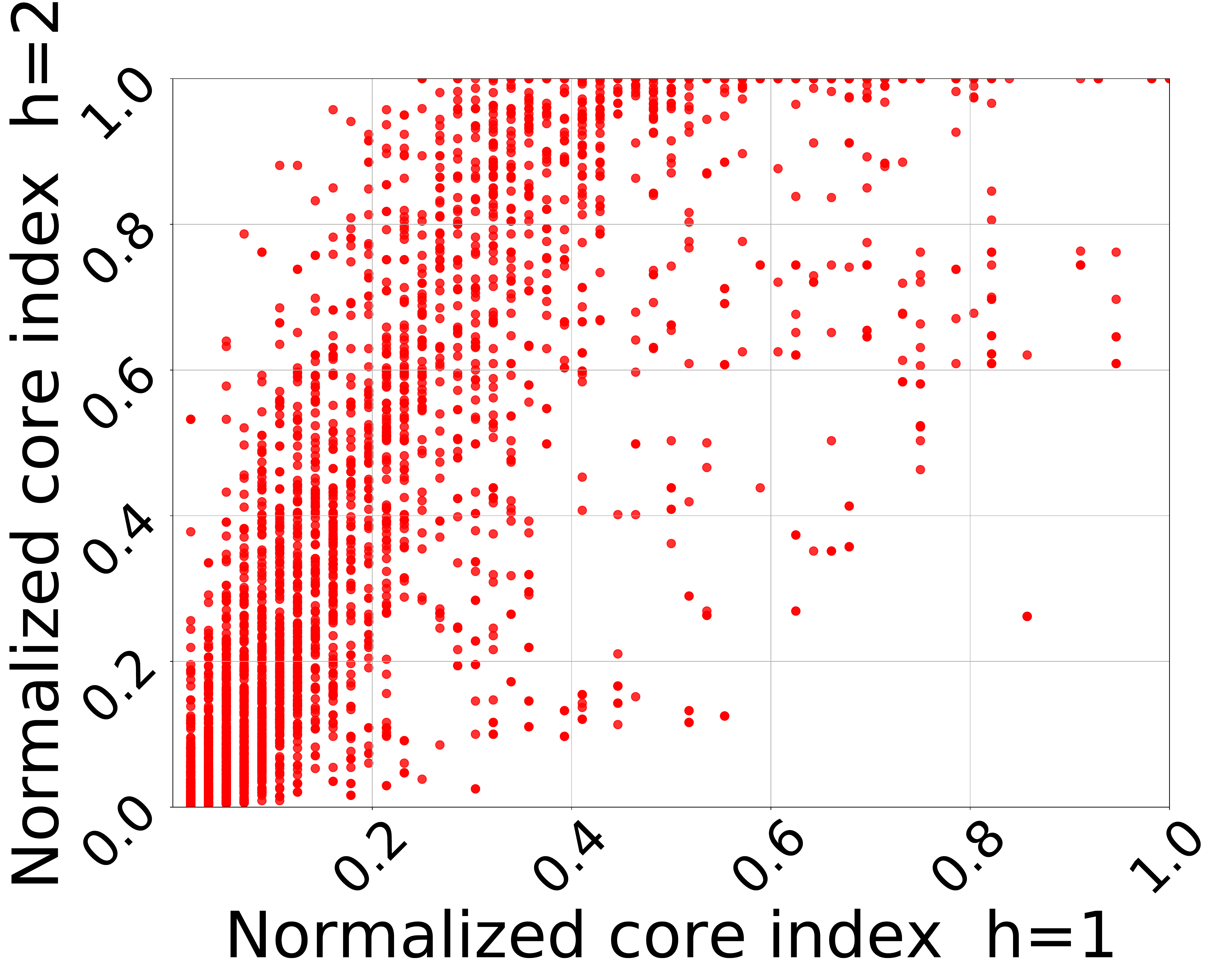}
{$h =1$ Vs. $h=2$}\label{Fig:ca-AstroPh:scatter2-5:2}
\end{minipage}\hfill
\begin{minipage}{0.23\textwidth}
\centering
\includegraphics[width=.95\linewidth]{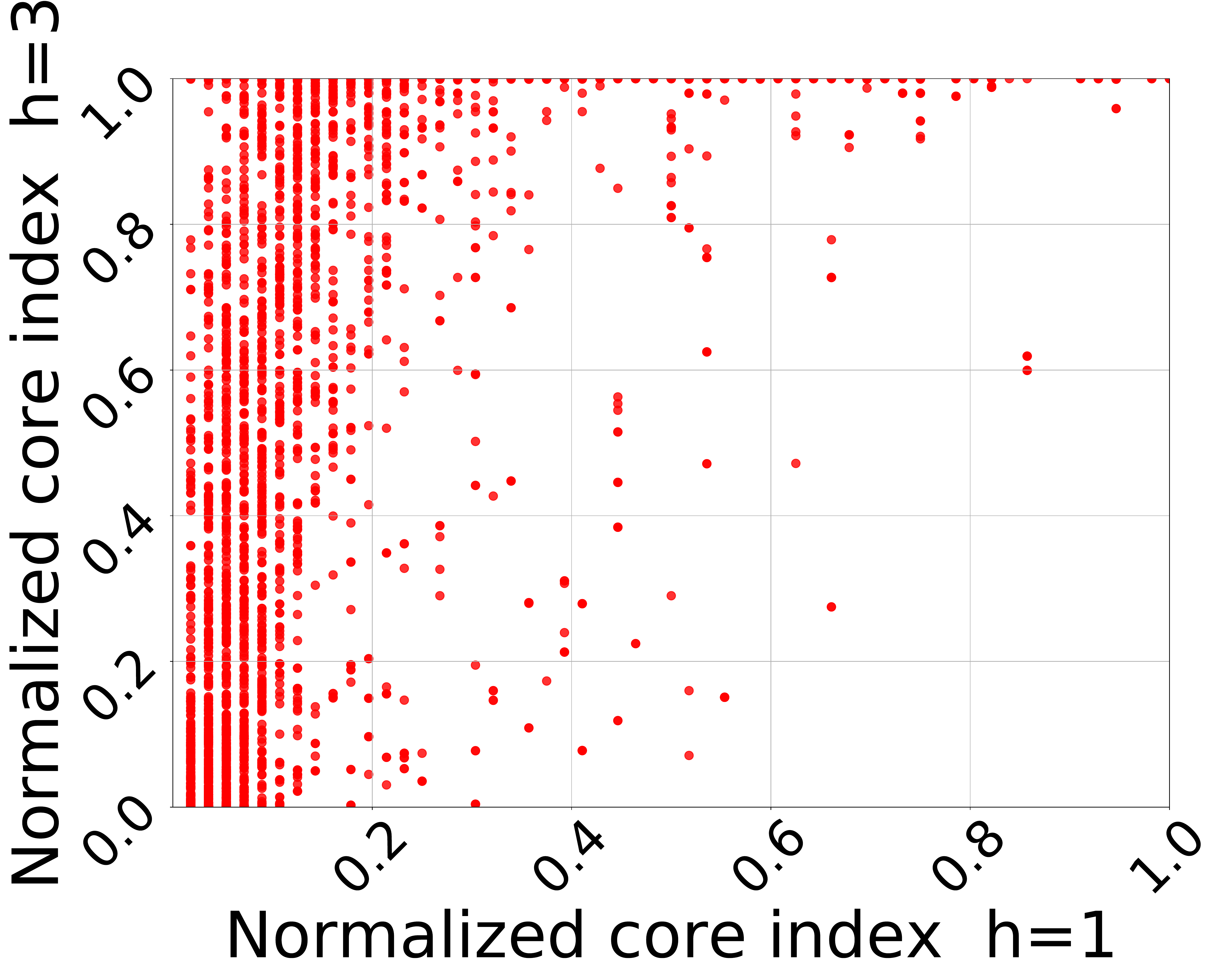}
{$h =1$ Vs. $h=3$}\label{Fig:ca-AstroPh:scatter2-5:3}
\end{minipage}

\smallskip

\begin{minipage}{0.23\textwidth}
\centering
\includegraphics[width=.95\linewidth]{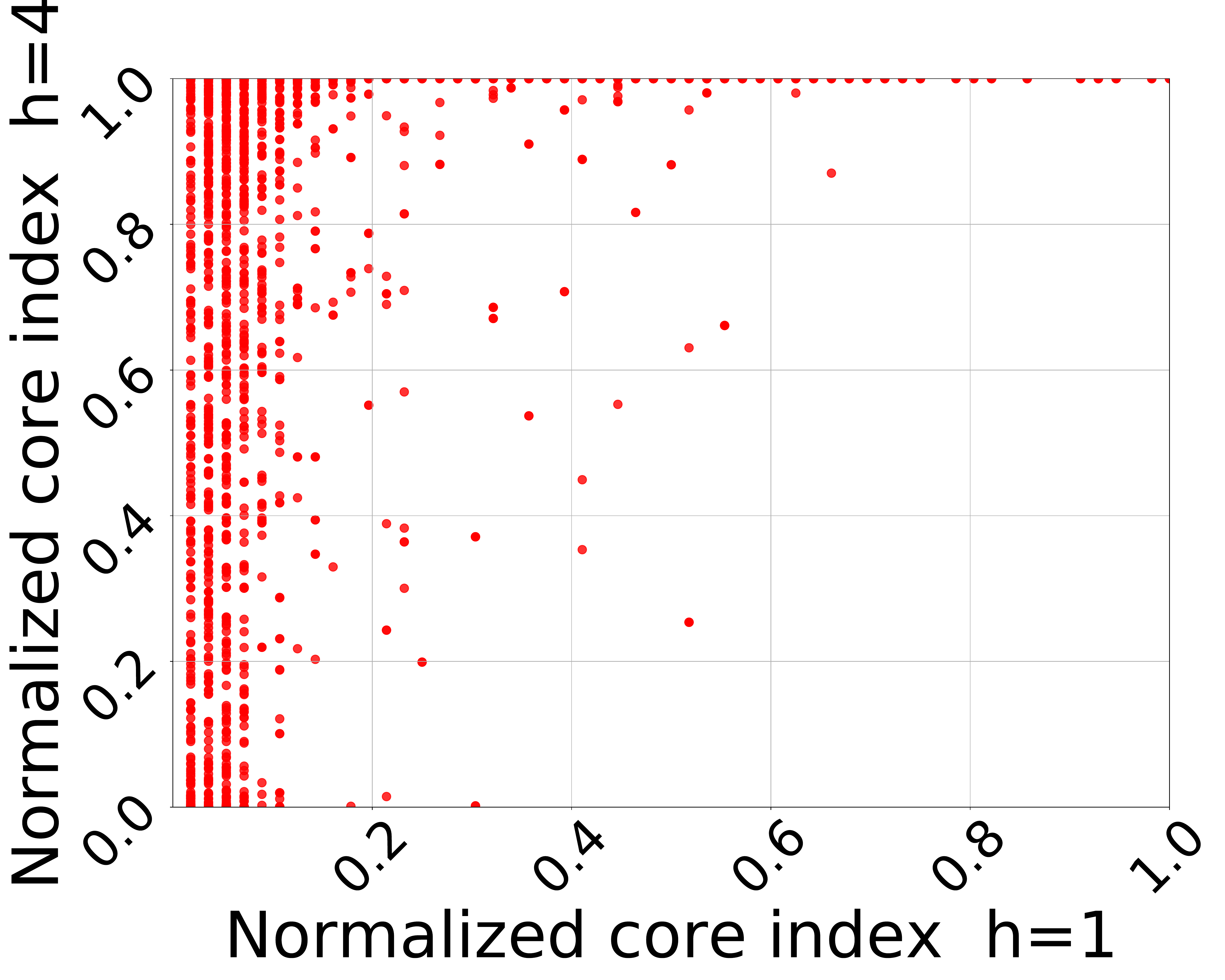}
{$h =1$ Vs. $h=4$}\label{Fig:ca-AstroPh:scatter2-5:4}
\end{minipage}
\begin{minipage}{0.23\textwidth}
\centering
\includegraphics[width=.95\linewidth]{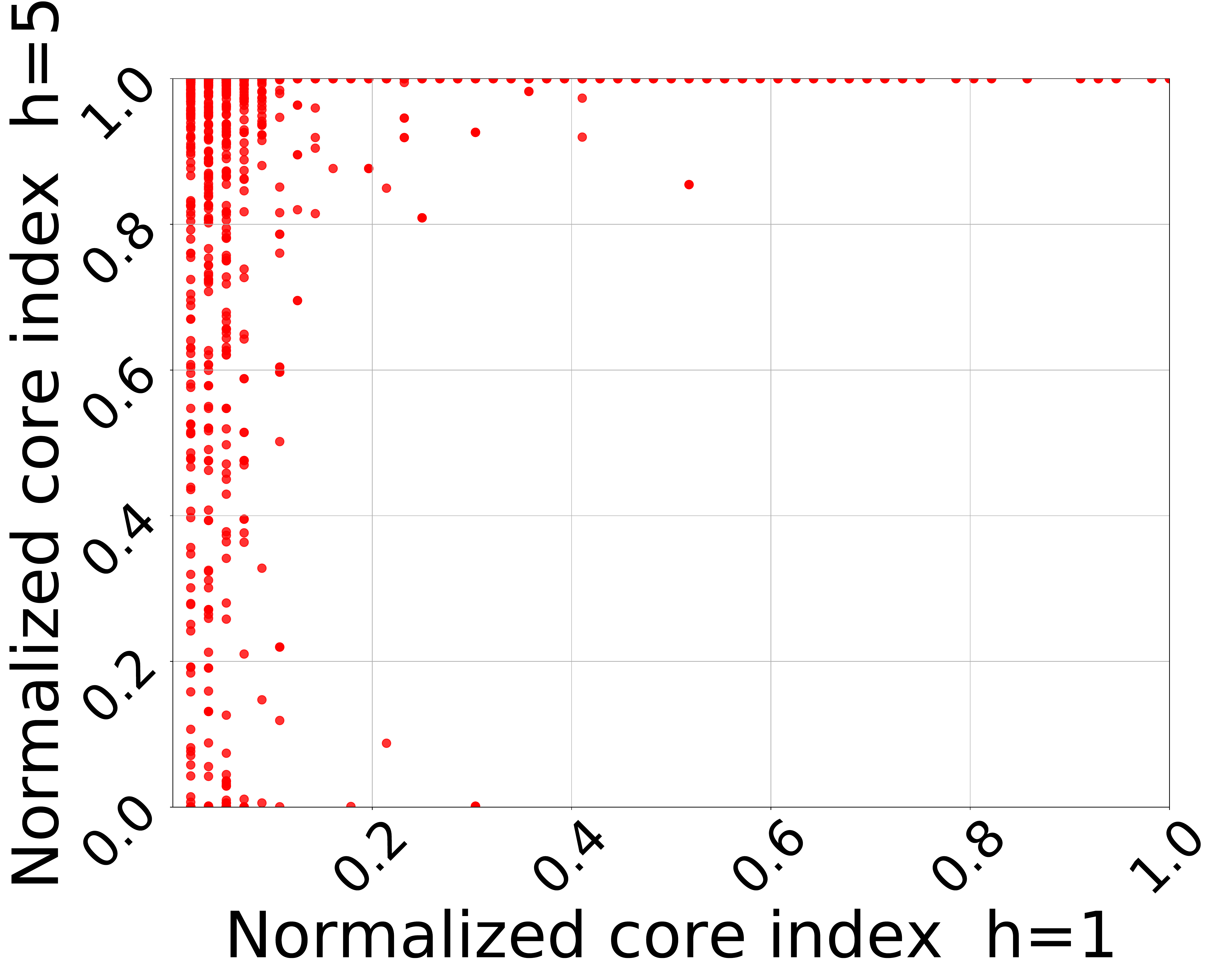}
{$h =1$ Vs. $h=5$}\label{Fig:ca-AstroPh:scatter2-5:5}
\end{minipage}
\vspace{-2mm}
\caption{Scatter plot of the core index of vertices for $h = 1$ Vs. core index  for $2\leq h\leq 5$ on \texttt{caAs}.}
\label{Fig:ca-AstroPh:scatter2-5}
\end{figure}

\spara{Characterization of the $(k,h)$-cores.}
Figure~\ref{Fig:ca-AstroPh:scatter2-5}  shows the diversity of information captured by different $h$: it reports scatter plots comparing, for 10\% of vertices randomly sampled, their core index for $h=1$ with their core index for $2\leq h\leq 5$. We can observe that there are vertices having normalized core index in $h=1$ above 0.6, and normalized core index in $h=3$ below 0.4. On the other hand, there are vertices with very low core index in $h=1$, and as $h$ grows can climb up in very high cores.
\begin{figure}[t!h!]
  \begin{minipage}{0.23\textwidth}
		\centering
		\includegraphics[width=.95\linewidth]{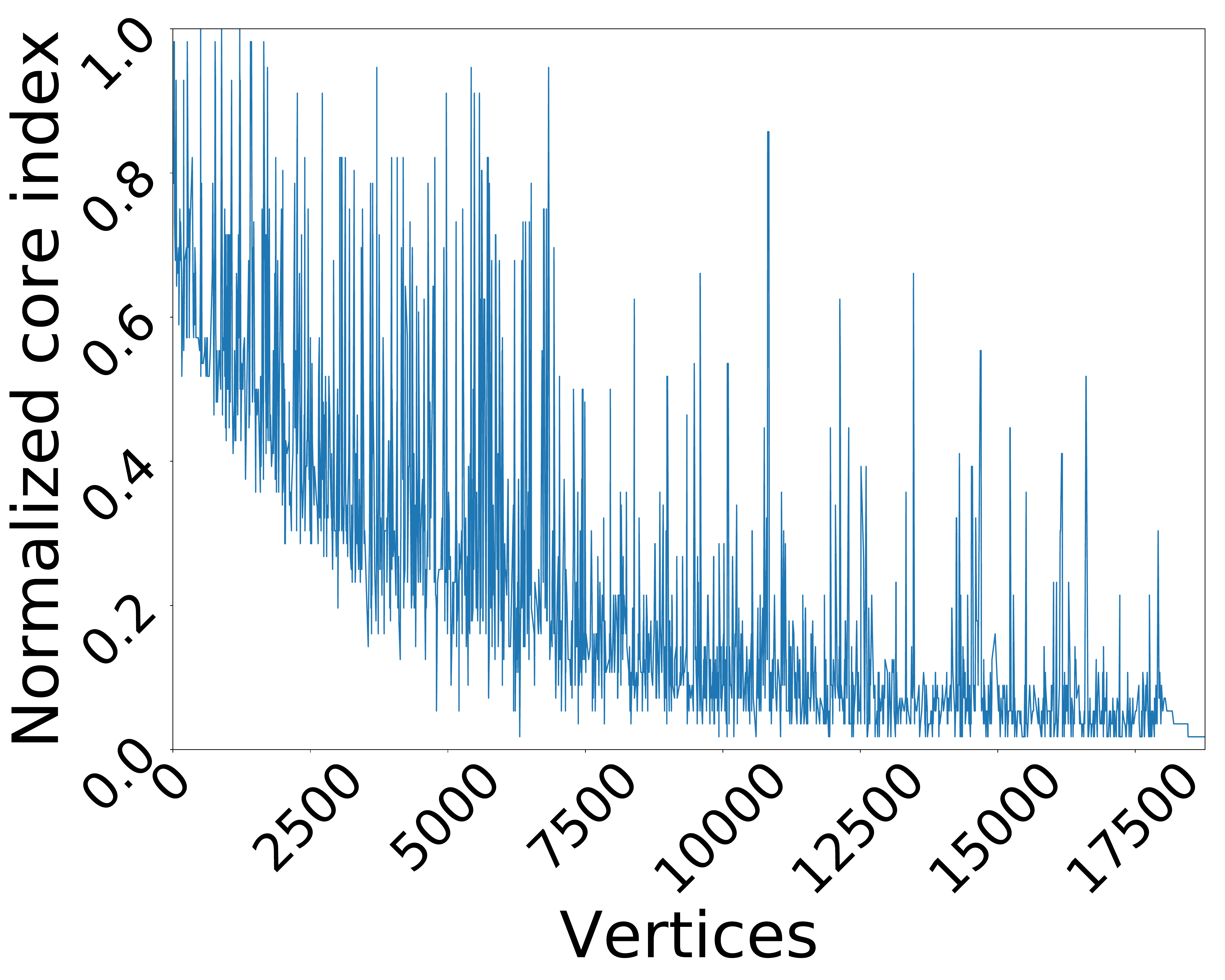}
		{$h=1$ }\label{Fig:ca-AstroPh:centralty1}
	\end{minipage}\hfill
	\begin{minipage}{0.23\textwidth}
		\centering
		\includegraphics[width=.95\linewidth]{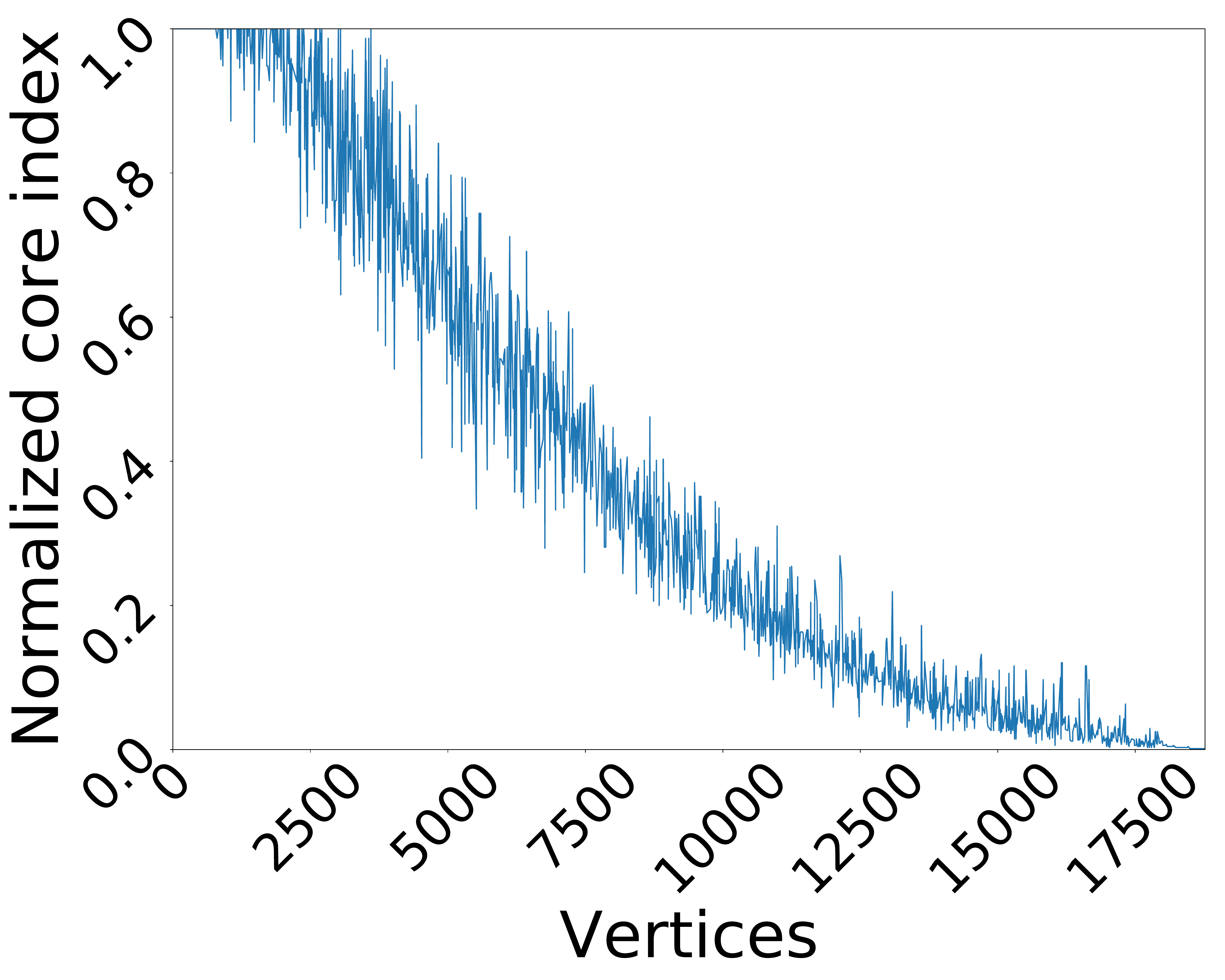}
		{$h=2$ }\label{Fig:ca-AstroPh:centralty2}
	\end{minipage}

\begin{minipage}{0.23\textwidth}
		\centering
		\includegraphics[width=.95\linewidth]{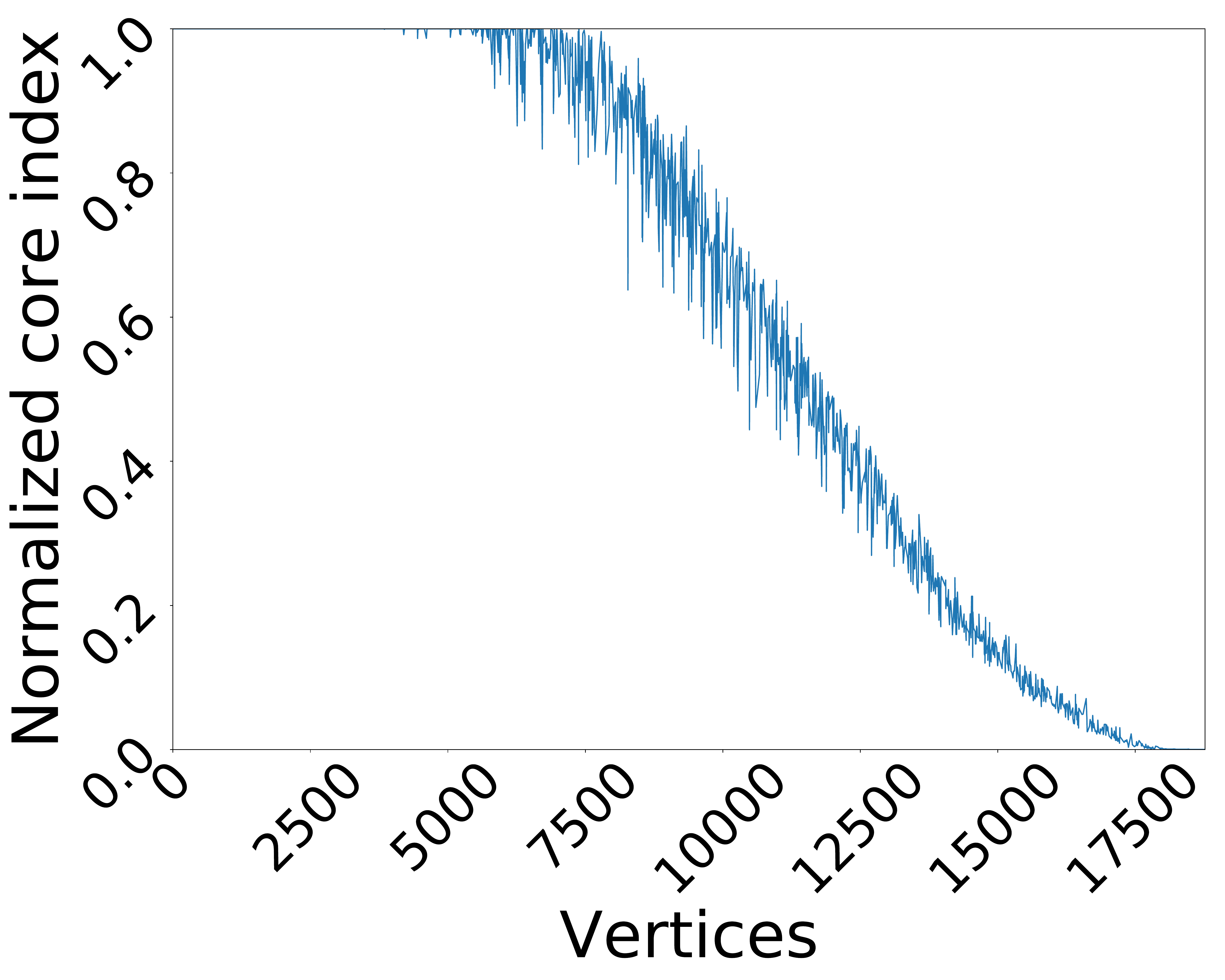}
		{$h=3$ }\label{Fig:ca-AstroPh:centralty3}
	\end{minipage}
	\begin{minipage}{0.23\textwidth}
		\centering
		\includegraphics[width=.95\linewidth]{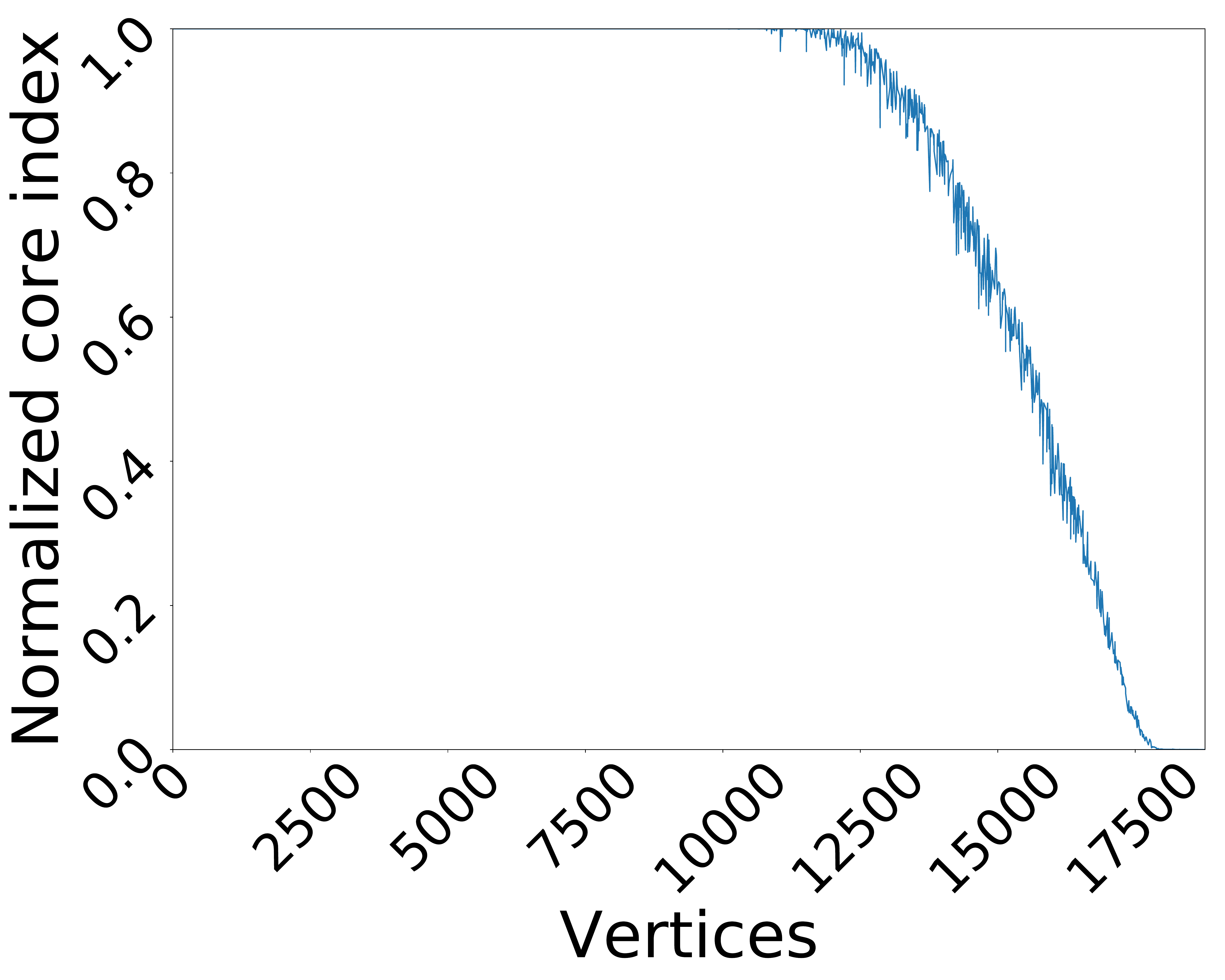}
		{$h=4$ }\label{Fig:ca-AstroPh:centralty4}
	\end{minipage}
\vspace{-2mm}
		\caption{Closeness centrality of vertices in \texttt{caAs}. The vertices on the $x$-axis are sorted in descending order of closeness centrality.}
			\label{Fig:ca-AstroPh:centrality}
\vspace{-2mm}
\end{figure}

Another feature of the distance-generalized core decomposition is that {\em as $h$ grows, the core index of a vertex is more correlated with its centrality}. This is to say that more central vertices end up belonging to higher cores.
In Figure~\ref{Fig:ca-AstroPh:centrality} we compare for each vertex its core index with its closeness centrality:
on the $x$-axis we present the vertices sorted in descending order of centrality (i.e., the closer to the origin, the more central it is), and on the $y$-axis we report their normalized core index. We can observe a stronger correlation as $h$ increases. In particular, for $h = 1$ we can have vertices which are not so central and which are in high cores, while for $h > 1$ this does not happen.

\section{Pseudo-code of Algorithm \ref{alg:h-club}}\label{appe:pc}

\begin{algorithm}[h!!]
\caption{Maximum $h$-Club Finding Algorithm}
\label{alg:h-club}
\begin{algorithmic}[1]
\REQUIRE graph $G=(V,E)$, distance threshold $h >1$, black-box algorithm $\mathcal{A}(G,h)$ for finding the maximum $h$-club in $G$
\ENSURE maximum $h$-club in $G$
\STATE perform $(k,h)$-core decomposition of $G$
\STATE $k_{cur}=k^*$ such that $C_{k^*}$ is the core of maximum index
\WHILE{maximum $h$-club not found}\label{alg:h-club:solve}
\STATE find maximum $h$-club in $G[ C_{k_{cur}}]$ via $\mathcal{A}(G[ C_{k_{cur}}],h)$
\IF{maximum $h$-club size in $G[ C_{k_{cur}}]>k_{cur}$}
\STATE maximum $h$-club found
\ELSE
\IF{maximum $h$-club size in $G[ C_{k_{cur}}]>0$}
\STATE $k_{cur}=min\{k_{cur}-1$, maximum $h$-club size\}
\ELSE
\STATE $k_{cur}=k_{cur}-1$
\ENDIF
\ENDIF
\ENDWHILE
\STATE return maximum $h$-club
\end{algorithmic}
\end{algorithm}

\begin{mdframed}[innerbottommargin=3pt,innertopmargin=3pt,innerleftmargin=6pt,innerrightmargin=6pt,backgroundcolor=gray!10,roundcorner=10pt]
\spara{Acknowledgments.} AK acknowledges support from MOE Tier-1 RG83/16 and NTU M4081678. FB and LS acknowledge support from Intesa Sanpaolo Innovation Center. The funders had no role in study design, data collection and analysis, decision to publish, or preparation of the manuscript.
\end{mdframed} 

\end{document}